\documentclass[10pt]{iopart}

\usepackage{cite}
\usepackage{color}

\usepackage[T1]{fontenc}
\usepackage[utf8]{inputenc}
\expandafter\let\csname equation*\endcsname\relax
\expandafter\let\csname endequation*\endcsname\relax
\usepackage{amsfonts}
\usepackage{amsbsy}
\usepackage{amsmath}
\usepackage{amssymb}
\usepackage{bbm}
\usepackage{bm}
\usepackage{epstopdf}
\usepackage{tikz}
\usetikzlibrary{arrows}
\newtheorem{Theorem}{Theorem}
\newtheorem{remark}{Remark}
\usepackage{physics}
\usepackage{url}

\begin{document}

\title[Quantum trajectories and output field properties for two-photon input field]{Quantum trajectories and output field properties for systems driven by two-photon input field}

\author{Anita D\k{a}browska$^1$, Gniewomir Sarbicki$^2$}

\address{$^1$Institute of Theoretical Physics and Astrophysics, University of Gda\'nsk, ul. Wita Stwosza 57, 80-308 Gda\'nsk, Poland,\\ $^2$Institute of Physics, Faculty of Physics, Astronomy and Informatics,  Nicolaus Copernicus University,
Grudzi\c{a}dzka 5/7, 87--100 Toru\'n, Poland}

\ead{anita.dabrowska@ug.edu.pl}

\vspace{10pt}

\begin{abstract}

The excitation of atomic and molecular systems by propagating light in a two-photon state within the Wigner-Weisskopf approximation has been described using stochastic tools. The problem of a stochastic evolution of the quantum system, depending on the results of the measurement of the output field, was formulated and solved making use of the model of repeated interactions and measurement. We defined the discrete in-time interaction between the quantum system and its environment being the electromagnetic field approximated by a chain or chains of harmonic oscillators. We determined analytical formulae for quantum trajectories associated with one-dimensional and two-dimensional counting processes, corresponding respectively to unidirectional or bidirectional input field prepared in the two-photon states. We derived the formulae for the exclusive probability densities of photon counts that allow us to completely characterize the photon statistics of the output field. 

Finally, we showed how to apply the quantum trajectories to obtain the formula for the probability of the two-photon absorption for a three-level atom in a ladder configuration. The paper also includes a discussion on the optimal two-photon state that maximizes the two-photon absorption probability.

\end{abstract}

%
%
%
%
%
\section{Introduction}

Non-classical states of propagating light \cite{BLPS90,Loudon2000,O06,RMS07} find applications in quantum communication \cite{Scarani09}, simulation \cite{Aaronon11}, metrology \cite{Giovannetti11}, and cryptography \cite{Zhong2022}. Together with the development of the methods of generating and manipulating such wave packets \cite{Banaszek05,Cooper13,Scarani2013,Lodahl15,Leong2016,Ogawa16,Lodahl17,Sedziak2017}, the theoretical descriptions of the scattering of non-classical light on quantum systems have been proposed. This problem was studied using the Heisenberg picture approach \cite{Domokos02,WMSS11,WMS12,Stolyarov2013,Scarani2016a},  the Lippmann-Schwinger equation \cite{Fan2009,Gritsev2012,Shen2015}, pure-state analysis \cite{Konyk16,Nysteen2015}, the input-output formalism and generalized master equations \cite{Fan2010,Gheri98,Baragiola12,Cirac2015,Molmer2019,Molmer2020}, and filtering equations 
\cite{Gough12a,Gough12b,Gough13,Dong15,Pan16,Zhang2016,Baragiola17,Dabrowska17,Dabrowska2019a,Dabrowska2019b,Dong19,Zhang2019a,Zhang2019b,Dabrowska2020,Gross2022,Dabrowska2023a,Dabrowska2023b}. In \cite{Fischer2018a, Fischer2018b} one can find the analysis referring to a discrete approximation of the bath Hilbert space. 

Significant effort is currently being devoted in quantum optics to developing methods for efficient excitation atoms and molecules by a propagating light containing a fixed number of photons. Particular interest is raised by using for the excitation of quantum systems by the two-photon light with photons entangled in frequencies (and time)  \cite{Schlawin2017,Schlawin2021,Raymer2021,Thew2021,Thew2022}. The inspiration and motivation for writing this paper come from research on the role of entanglement and interference in two-photon absorption. Our results published in \cite{Dabrowska17,Dabrowska2019b,Dabrowska2023a} concerned only unidirectional light carrying photons with identical temporal profiles and they considered pulse of a single central frequency. The previous studies did not address more general light states in which photons could be entangled in frequency (and time). Our objective is to generalize these results first to the case where the photons can differ in temporal profiles and in central frequencies and then to arbitrary pure two-photon state. Referring to laboratory practice, using the spontaneous parametric down-conversion method to generate entangled photon light, we consider excitation of the quantum system by unidirectional and bidirectional two-photon light. We describe the interaction of light with a quantum system using the Wigner-Weisskopf approximation. Thus we consider the model formulated in the framework of standard assumptions made in quantum optics such as  a flat coupling constant, rotating wave-approximation, and the extension of the lower limit of integration over frequency to minus infinity  \cite{WM10, Scully1997}. Thus the bandwidth of the spectrum is assumed to be much smaller than the central frequency of the pulse. 

We solve the problem of driving a quantum system by the two-photon light using quantum filtering theory \cite{BarBel91,Car93,GZ10,WM10,HP84,Par92} that is formulated within the framework of the input-output formalism \cite{GZ10}. The filtering equations describe the evolution of a quantum system that depends on the results of continuous observations of the output field, i.e., the field after the interaction with the quantum system. The measurement results are in general random and hence we are dealing here with stochastic evolution of the system. The form of the filtering equations depends on the state of the input field and the type of measurement and solutions to them are called a posteriori states or quantum trajectories. The descriptions of the stochastic evolution of the open system for non-classical fields have appeared relatively recently. In the literature there exist derivations of the filtering equations based on cascaded systems \cite{Gough12a}, non-Markovian embedding  \cite{Gough12b,Gough13,Zhang2016,Zhang2019b}, and the input field decomposition \cite{Baragiola17,Gross2022}. As shown in \cite{Dabrowska17,Dabrowska2019a,Dabrowska2019b,Dabrowska2023a,Dabrowska2023b}, quantum trajectories for the non-classical state of the input field, can also be effectively determined employing a model of discrete 
in time of repeated interactions (collision model) \cite{B02,GS04,AP06,PP09, P10,Ciccarello2017,Gross18,Ciccarello2021,Filippov2022}. A mathematically rigorous formulation of the discrete model of filtration can be found in \cite{GS04,AP06,P10}.  An introduction to the collision model in quantum optics is given in \cite{Ciccarello2017}. The interaction of a quantum system with propagating light containing a fixed number of photons is inherently complex. Stochastic methods provide a framework for analyzing both the reduced evolution of the open system and the properties of the output field.

In this paper, we show how to determine quantum trajectories and how to use them to find the unconditional evolution of an open system and the photon statistics of the output field for the input field prepared in the two-photon vector state. We consider in this paper a collision model with the electromagnetic field approximated by one or respectively by two chains of non-interacting harmonic oscillators that sequentially interact with the quantum system. We assume that the environment of the system is prepared in an entangled state being a discrete analogue of a continuous-mode two-photon state. The entanglement of the field harmonic oscillators makes the evolution of the open system non-Markovian \cite{Ciccarello2018,Dabrowska2021}. We discuss the discrete stochastic dynamics of an open system that depend on the results of measurement performed on the output field and determine analytical formulae for the quantum trajectories associated with the photon counting process, describe their general structure, and provide the physical interpretation to them. Starting from a discrete-in-time description, we eventually obtain expressions for the evolution depending on the results of the continuous-in-time observation of the output field. The paper includes also formulae for the exclusive probability densities of photon counts for one- and two-dimensional counting processes. The collision-model-based approach is intuitive and facilitates understanding of temporal correlations in a field with a fixed number of photons, as well as the entanglement of the system with both the input and output fields, resulting in the system's non-Markovian evolution.  An important issue from a mathematical perspective is that the proposed method for solving the problem of the interaction between a quantum system and two-photon light using conditional vectors reduces the number of equations. Instead of $16$ equations for stochastic operators, as shown in \cite{Dong15,Zhang2016,Dong19,Zhang2019b}, the set derived in this paper consists of $4$ coupled stochastic equations; both in the case of unidirectional and bidirectional fields.

In the second part of the paper we show how to apply quantum trajectories to determine the analytical formula for the probability of two-photon absorption. We consider here a three-level atom in a ladder configuration.  We derive formulae for the probability of two-photon absorption for an arbitrary pure two-photon state of light. We would like to stress that our results are strict within the Wigner-Weisskopf approximation. In this paper, we present, moreover, a discussion on the optimal two-photon state giving the maximum achievable value of the two-photon absorption for unidirectional and bidirectional fields. We consider two different cases of system excitation by a two-photon field with indistinguishable and distinguishable photons, highlighting the role of resonances and entanglement in the optimal excitation of the system. The selected example has fundamental importance in the discussion on the use of two-photon light for exciting atomic systems.

The paper is organized as follows. In sections 2, 3, and 4, we consider the interaction of a quantum system with a unidirectional field. In section 2, we formulate the model of repeated interactions for such an environment. In section 3, we define the model of repeated measurements and determine the formula for the a posteriori state of the open system. In section 4, we derive the formulae for quantum trajectories, describe their general structure, and demonstrate how to use them to find the statistics of photons in the output field. Sections 5, 6, and 7 present analogous considerations for the bidirectional field. Section 8 is devoted to the two-photon absorption problem for a three-level atom in a ladder configuration. Appendices provide proofs and additional calculations.

\section{Repeated interactions model for unidirectional field} 

In this section, we describe the basic properties of the model of discrete interactions between a quantum system $\mathcal{S}$ of the Hilbert space $\mathcal{H_{S}}$ and an environment $\mathcal{E}$ being an unidirectional electromagnetic field approximated by a sequence of $M$ harmonic oscillators. We will carry out our analysis at a general level, without specifying the system 
$\mathcal{S}$. As usual in stochastic evolution analyses, we describe the propageting electromagnetic field in time domain. The Hilbert space of the environment is given by
\begin{equation}
\mathcal{H}_{\mathcal{E}}=\bigotimes_{k=0}^{M-1}\mathcal{H}_{\mathcal{E},k},
\end{equation}
where $\mathcal{H}_{\mathcal{E},k}$ is the Hilbert space of the harmonic oscillator which interacts with $\mathcal{S}$ in the time interval $[k\tau, (k+1)\tau)$.
For brevity, we will use the notation $\mathcal{H}_{\mathcal{E}}^{j-1]}=\bigotimes_{k=0}^{j-1}\mathcal{H}_{\mathcal{E},k}$ (the past environment space) and 
$\mathcal{H}_{\mathcal{E}}^{[j}=\bigotimes_{k=j}^{M-1}\mathcal{H}_{\mathcal{E},k}$ (the future environment space). The terms ``past'' and ``future'' refer to the field before and after the interaction with the system $\mathcal{S}$, respectively. The composed system, environment and system $\mathcal{S}$, is described in the Hilbert space $\mathcal{H}_{\mathcal{E}}\otimes \mathcal{H}_{\mathcal{S}}.$
We assume that the field harmonic oscillators do not interact with each other but they interact one by one with the system $\mathcal{S}$.  Each interaction (``collision'') lasts for the time $\tau$.  
The discrete time evolution of the composed system $\mathcal{E}+\mathcal{S}$ from time zero to time $\tau j $ for $1\leq j\leq M-1$ is defined as \cite{AP06,PP09, P10,Ciccarello2017,Fischer2018a,Gross18}
\begin{equation}\label{disc_evol}
\hat{U}_{j} = \hat{\mathbb{V}}_{j-1} \hat{\mathbb{V}}_{j-2} \ldots \hat{\mathbb{V}}_{0},\;\;\;\;\;\hat{U}_{0}=\hat{\mathbbm{1}},
\end{equation}
where a unitary operator $\hat{\mathbb{V}}_{k}$ describing the interaction of $\mathcal{S}$ with the field in the interval $[k\tau, (k+1)\tau)$
is defined as 
\begin{equation}\label{intermat}
\hat{\mathbb{V}}_{k}= \hat{\mathbbm{1}}_\mathcal{E}^{k-1]} \otimes \hat{\mathbb{V}}_{[k}  ,
\end{equation}
where
\begin{equation}\label{Vk}
\hat{\mathbb{V}}_{[k} =  \exp\left(-i\tau \hat{H}_{k}\right)
\end{equation}
and
\begin{equation}\label{hamint}
\hat{H}_{k} = \hat{\mathbbm{1}}_{\mathcal{E}}^{[k}\otimes \hat{H}_{\mathcal{S}}+\frac{i}{\sqrt{\tau}}\left(\hat{b}_{k}^{\dagger}\otimes\hat{\mathbbm{1}}_{\mathcal{E}}^{[k+1}\otimes \hat{L}-\hat{b}_{k}\otimes\hat{\mathbbm{1}}_{\mathcal{E}}^{[k+1}\otimes \hat{L}^{\dagger}\right).
\end{equation}
Note that the operator $\hat{\mathbb{V}}_{[k}$ acts on the space $\mathcal{H}_{\mathcal{E}}^{[k}\otimes \mathcal{H_{S}}$ but it has nontrivial action only on $\mathcal{H}_{\mathcal{E},k}\otimes \mathcal{H_{S}}$.
For simplicity, we set the Planck constant $\hbar=1$. The evolution of the composed system $\mathcal{E}+\mathcal{S}$ is written in the interaction picture with respect to the free dynamics of the field. Here $\hat{H}_{\mathcal{S}}$ stands for the Hamiltonian of the system $\mathcal{S}$ and $\hat{L} \in \mathcal{B(H_{S})}$, i.e., it is a linear bounded operator acting from $\mathcal{H_{S}}$ into $\mathcal{H_{S}}$. We denote by $\hat{b}_{k}$ and $\hat{b}_{k}^{\dagger}$, respectively, the annihilation and creation operators of the harmonic oscillator of number $k$, so they act 
 as 
 \begin{equation}
\hat{b}_{k}\ket{n}_{k}=\sqrt{n}\ket{n-1}_{k},
\end{equation}
\begin{equation}
\hat{b}_{k}^{\dagger}\ket{n}_{k}=\sqrt{(n+1)}\ket{n+1}_{k},
\end{equation}
where $\ket{n}_{k}$ is the number state in the space $\mathcal{H}_{\mathcal{E},k}$. 
The operators $\hat{b}_{k}$ and $\hat{b}_{l}^{\dagger}$ satisfy the standard canonical commutation relations:
\begin{equation}\label{CCR1}
[\hat{b}_{k}, \hat{b}_{l}] = 0,\;\;\; [\hat{b}_{k}^{\dagger}, \hat{b}_{l}^{\dagger}] = 0,\;\;\;
[\hat{b}_{k}, \hat{b}_{l}^{\dagger}] =\delta_{kl}. 
\end{equation}
Note that the operator $\hat{U}_{j}$ acts non-trivially in the past space $\mathcal{H}_{\mathcal{E}}^{j-1]}$, i.e., in the Hilbert space associated with the harmonic oscillators that have already interacted with $\mathcal{S}$ and $\hat{U}_{j}$ acts trivially in the future space $\mathcal{H}_{\mathcal{E}}^{[j}$ which refers to the harmonic oscillators that will interact with $\mathcal{S}$ after $j\tau$. 
Clearly, the input field is formed by the sequence of harmonic oscillators before the interaction with the system $\mathcal{S}$, while the output field by the harmonic oscillators after this interaction.

To define the Fock state in the Hilbert space $\mathcal{H}_{\mathcal{E}}$, let us first introduce in $\mathcal{H}_{\mathcal{E}}^{[j}$ the creation operator,
\begin{equation}\label{creation}
\hat{B}_{[j}^{\dagger}[\xi]=\sum_{k=j}^{M-1}\sqrt{\tau}\xi_{k}\hat{\mathbbm{1}}_\mathcal{E}^{[j,k-1]} \otimes \hat{b}_{k}^{\dagger}\otimes\hat{\mathbbm{1}}_\mathcal{E}^{[k+1},
\end{equation}
where 
${\xi}_{k}\in \mathbb{C}$ and 
$\sum_{k=0}^{M-1} \tau\lvert {\xi}_{k}\rvert^2=1$. 
One can check that the commutator of $\hat{B}_{[j}^{\dagger}[\xi]$ and its Hermitian-conjugate operator $\hat{B}_{[j}[\xi]$ has the form 
\begin{equation}
[\hat{B}_{[j}[\xi],\hat{B}_{[j}^{\dagger}[\xi]]=\sum_{k=j}^{M-1}\tau \lvert\xi_{k}\rvert^2.
\end{equation}
One can use the creation operator (\ref{creation}) to construct the number vectors in the Hilbert space $\mathcal{H}_{\mathcal{E}}^{[j}$ by the formula
\begin{equation}\label{vectorm}
\ket{n_{\xi}}_{[j} = \frac{1}{\sqrt{n!}}\left(\hat{B}_{[j}^{\dagger}[\xi]\right)^{n}\ket{vac}_{[j},
\end{equation}
where $n$ stands for a photon number and $\ket{vac}_{[j} =\ket{0}_{j}\otimes\ket{0}_{j+1}\otimes\ldots\ket{0}_{M-1}$ is the vacuum vector in $\mathcal{H}_{\mathcal{E}}^{[j}$. 
Notice, that in contrary to the formula (\ref{vil}), where single-time number states appear, the ket index is now multi-temporal, and the photon temporal shape is indicated by $\xi$. 
We briefly outline the properties of these vectors.
One can check that the number vectors are mutually orthogonal: 
\begin{equation}
_{[j}\bra{n^{\prime}_{\xi}}\ket{n_{\phi}}_{[j}=\delta_{n^{\prime}n}\left(\sum_{k=j}^{M-1}\tau\xi_{k}^{\ast}\phi_{k}\right)^n,
\end{equation}
where ${\phi}_{k}\in \mathbb{C}$ and 
$\sum_{k=0}^{M-1} \tau\lvert {\phi}_{k}\rvert^2=1$. Furthermore, acting with the operator $\hat{B}_{[j}^{\dagger}[\xi]$ on the number vector one obtains
\begin{equation}\label{prop1}
\hat{B}_{[j}^{\dagger}[\xi]\ket{n_{\xi}}_{[j} =\sqrt{n+1} \ket{(n+1)_{\xi}}_{[j}
\end{equation}
and with the annihilation operator, one gets
\begin{equation}\label{prop3}
\hat{B}_{[j}[\xi]\ket{n_{\xi}}_{[j} =\sqrt{n} \sum_{k=j}^{M-1}\tau\lvert\xi_{k}\rvert^2\ket{(n-1)_{\xi}}_{[j}. 
\end{equation}
The important property of the number vectors that we shall use in the next section is the additive decomposition \cite{Dabrowska2019b}, 
\begin{equation}\label{adp}
\ket{n_{\xi}}_{[j}=\sum_{n^{\prime}=0}^{n}
\sqrt{\binom{n}{n^{\prime}}}
(\sqrt{\tau}\xi_{j})^{n^{\prime}}\ket{n^{\prime}}_{j}\otimes \ket{(n-n^{\prime})_{\xi}}_{[j+1}. 
\end{equation}
Let us notice, moreover, that the form of the formula (\ref{adp}) clearly shows that we are dealing with an entangled vector of the harmonic oscillators in $\mathcal{H}_{\mathcal{E}}^{[j}$.

We shall consider in this paper the two-photon vector state in $\mathcal{H}_{\mathcal{E}}$ defined by 
\begin{equation}\label{2phot}
\ket{2_{\xi\phi}} = \frac{1}{\sqrt{\mathcal{N}_{\xi\phi}}}\hat{B}^{\dagger}[\xi]\hat{B}^{\dagger}[\phi]\ket{vac},
\end{equation}
where $\hat{B}^{\dagger}[\xi]=\hat{B}_{[0}^{\dagger}[\xi]$, $\hat{B}^{\dagger}[\phi]=\hat{B}_{[0}^{\dagger}[\phi]$,  $\ket{vac}=\ket{0}_{0}\otimes\ket{0}_{1}\otimes\ldots\ket{0}_{M-1}$ is the vacuum vector in $\mathcal{H}_{\mathcal{E}}$, and 
$\mathcal{N}_{\xi\phi} = \braket{2_{\xi\phi}}$ 
is the normalization factor.
In general, state (\ref{2phot}) may be a state in which the photons are correlated \cite{Benatti2020}.
Note that the state (\ref{2phot}) is a discrete analogue of the continuous-mode number state \cite{BLPS90,Loudon2000,O06,RMS07}. 

The definition of (\ref{2phot}) can be generalized. Let us observe that the formula (\ref{2phot}) is linear in $\phi$ and $\xi \in \mathbb{C}^M$ and hence have a unique continuation to a linear map from $\mathbb{C}^M \otimes \mathbb{C}^M$ to the space of two-photon (unormalised) state vectors $\mathrm{span}\{\hat{\tilde b}_i^\dagger \hat{\tilde b}_j^\dagger \ket{vac}, 0 \le i \le j < M \} \subset \mathcal{H}_{\mathcal{E}}$: 
\begin{equation} \label{isom}
\mathfrak{F}: \Phi \in \mathbb{C}^M \otimes \mathbb{C}^M \mapsto \ket{2_{\Phi}} = \sum_{k_1, k_2 =0}^{M-1} \tau \Phi_{k_2, k_1} \hat{\tilde b}_{k_2}^\dagger \hat{\tilde b}_{k_1}^\dagger \ket{vac},
\end{equation}
where
\begin{equation}
\hat{\tilde{b}}_{k}^{\dagger}=\mathbbm{\hat{1}}_\mathcal{E}^{k-1]} \otimes \hat{b}_{k}^{\dagger}\otimes\mathbbm{\hat{1}}_\mathcal{E}^{[k+1}.
\end{equation}
Having two-photon states $\ket{2_{\Phi}}$ and $\ket{2_{\Psi}}$, one can express their inner product by the standard inner product $\circ$ 
in $\mathbb{C}^M \otimes \mathbb{C}^M$ :
\begin{equation} \label{inner_prod}
    \bra{2_{\Phi}}\ket{2_{\Psi}} = 
    \Phi \circ (I+\mathcal{F}) \Psi = 2 \Phi \circ \Pi_\mathcal{S} \Psi,
\end{equation}
where $\mathcal{F}$ denotes the flip operator and $\Pi_\mathcal{S}$ the projector onto symmetric subspace.
On the symmetric subspace one has $\bra{2_{\Phi}}\ket{2_{\Psi}} = 2 \Phi \circ \Psi$ and hence $\frac 12 \mathfrak{F}$ establishes an isometry (Hilbert spaces isomorphism) between symmetric subspace of $\mathbb{C}^M \otimes \mathbb{C}^M$ and the space of two-photon states. Let us notice further that the antisymmetric subspace of $\mathbb{C}^M \otimes \mathbb{C}^M$ is the kernel of $\mathfrak{F}$ - the antisymmetric part of $\Phi \in \mathbb{C}^M \otimes \mathbb{C}^M$ gives no contribution to the state $\ket{2_\Phi}$.  The formula (\ref{isom}) constitutes an isomorphism between the symmetric subspace of $\mathbb{C}^M \otimes \mathbb{C}^M$ and the space of two-photon states. As a direct consequence of (\ref{inner_prod}) one calculates the normalisation factor:
\begin{equation}\label{norm}
    \mathcal{N}_{\Phi} = \braket{2_\Phi} = \sum_{k_2,k_1=0}^{M-1}\tau^2 |\Phi_{k_2,k_1}|^2 
    + \sum_{k_2,k_1=0}^{M-1} \tau^2\Phi^*_{k_1,k_2} \Phi_{k_2,k_1},
\end{equation}
in particular:
\begin{equation}
    \mathcal{N}_{\xi\phi} = \braket{2_{\xi\phi}} = 1 + \left|\sum_{k=0}^{M-1}\tau\xi_{k}^{\ast}\phi_{k}\right|^2,
\end{equation}
Since now we will consider only normalised two-photon states ($\braket{2_\Phi} = 1$). 

We assume the composed system $\mathcal{E}+\mathcal{S}$ is prepared initially in the product state 
\begin{equation}\label{ini0}
\ket{\Psi_{0}}=\ket{2_{\xi\phi}}\otimes\ket{\psi_{0}},
\end{equation}
where $\ket{2_{\xi\phi}}$ is the two-photon vector state defined by (\ref{2phot}) and $\ket{\psi_{0}}$ is the initial state of $\mathcal{S}$.

\section{Repeated measurements and conditional state for unidirectional field}

This section is devoted to presenting the formalism of repeated measurements carried out on the output field. Let us note that the output field carries information about the system $\mathcal{S}$. In this section, we describe a conditional evolution of the composed system consisting of the input field and $\mathcal{S}$. Let us suppose that after each interaction the measurement is performed on the output field, to be more precise, on the harmonic oscillators just after their interaction with the system $\mathcal{S}$.  Clearly, the conditional evolution depends on the type of measurement performed on the output field. In this paper we shall study the evolution conditioned by the results of the measurement of the observable
\begin{equation}\label{obs}
\hat{b}_{k}^{\dagger}\hat{b}_{k}, \;\;\; k=0,1,\ldots, M-1.
\end{equation}
Considering a photon counting measurement, it is convenient to use for the operator (\ref{Vk}) the photon number  representation, i.e.,   
\begin{equation}\label{vil}
\exp\left(-i\tau \hat{H}_{k}\right)  = \sum_{nn^{\prime}} |n\rangle_{k} {}_{k}\langle n^{\prime}|\otimes\hat{\mathbbm{1}}_{\mathcal{E}}^{[k+1} \otimes \hat{V}_{nn^{\prime}},
\end{equation}
where $\hat{V}_{nn^{\prime}} \in\mathcal{B}(\mathcal{H}_S)$ and $n,n^{\prime}=0,1,2,\ldots$. The explicit form of $\hat{V}_{nn^{\prime}}$ for $n, n^{\prime}=0,1$ are given in (\ref{vmatrix}). 
We represent the results of measurements performed up to time $j\tau$ by the stochastic vector $\pmb{\eta}_j=(\eta_j,\eta_{j-1},\ldots,\eta_1)$. Our analysis can be summarised by the following theorem.

\begin{Theorem}\label{TH-1} The conditional state vector of the input part of the environment (the part of the environment which has not interacted with $\mathcal{S}$ up to $j\tau$) and the system $\mathcal{S}$, for the initial state (\ref{ini0}) and the measurement of (\ref{obs}), at time $j\tau$ is given by
	\begin{equation}\label{cond1}
	\ket{\tilde{\Psi}_{j}} 
	= \frac{\ket{\Psi_{j}}}
	{\sqrt{\bra{\Psi_{j}}\ket{\Psi_{j}}}},
	\end{equation}
	where $\ket{\Psi_{j}}$ is the unnormalized conditional vector from the Hilbert space $\mathcal{H}_\mathcal{E}^{[j} \otimes \mathcal{H}_\mathcal{S}$ having the form
	\begin{equation}\label{condTH1}
	\ket{\Psi_{j}}=
	\ket{vac}_{[j}\otimes\ket{\psi_{j}(0)}+
	\ket{1_{\xi}}_{[j}\otimes\ket{\psi_{j}(1_{\xi})}+\ket{1_{\phi}}_{[j}\otimes\ket{\psi_{j}(1_{\phi})}+\ket{2_{\xi\phi}}_{[j}\otimes\ket{\psi_{j}(2_{\xi\phi})},
	\end{equation}
	where $\ket{2_{\xi\phi}}_{[j}= \frac{1}{\sqrt{\mathcal{N}_{\xi\phi}}}\hat{B}^{\dagger}_{[j}[\phi]\hat{B}^{\dagger}_{[j}[\xi]\ket{vac}_{[j}$ and the conditional vectors $\ket{\psi_{j}(0)}$, $\ket{\psi_{j}(1_{\xi})}$, $\ket{\psi_{j}(1_{\phi})}$, $\ket{\psi_{j}(2_{\xi\phi})}$ from the Hilbert space $\mathcal{H}_{\mathcal{S}}$ satisfy the set of coupled recurrence equations:
	\begin{align}\label{TH-1_rec1}
	\ket{\psi_{j+1}(0)} =& \hat{V}_{\eta_{j+1}0}\ket{\psi_{j}(0)}+\sqrt{\tau}\xi_{j}\hat{V}_{\eta_{j+1}1}\ket{\psi_{j}(1_{\xi})}+\sqrt{\tau}\phi_{j}\hat{V}_{\eta_{j+1}1}\ket{\psi_{j}(1_{\phi})}\nonumber\\
 &+\frac{\sqrt{2}\tau\xi_{j}\phi_{j}}{\sqrt{\mathcal{N}_{\xi\phi}}}\hat{V}_{\eta_{j+1}2}\ket{\psi_{j}(2_{\xi\phi})},\\
	\ket{\psi_{j+1}(1_{\xi})} =& \hat{V}_{\eta_{j+1}0}\ket{\psi_{j}(1_{\xi})}+\frac{1}{\sqrt{\mathcal{N}_{\xi\phi}}}\sqrt{\tau}\phi_{j}\hat{V}_{\eta_{j+1}1}\ket{\psi_{j}(2_{\xi\phi})},\\
	\ket{\psi_{j+1}(1_{\phi})} =&\hat{V}_{\eta_{j+1}0}\ket{\psi_{j}(1_{\phi})}+\frac{1}{\sqrt{\mathcal{N}_{\xi\phi}}}\sqrt{\tau}\xi_{j}\hat{V}_{\eta_{j+1}1}\ket{\psi_{j}(2_{\xi\phi})}
	,\\
	\ket{\psi_{j+1}(2_{\xi\phi})} =& \hat{V}_{\eta_{j+1}0}\ket{\psi_{j}(2_{\xi\phi})},\label{TH-1_rec2}
	\end{align}
and have the initial values $\ket{\psi_{j=0}(0)}=\ket{\psi_{j=0}(1_{\xi})}=\ket{\psi_{j=0}(1_{\phi})}=0$, $\ket{\psi_{j=0}(2_{\xi\phi})}=\ket{\psi_{0}}$.
	\end{Theorem}
For the proof of Theorem \ref{TH-1}, see \ref{ProofTH-1}. We obtained the formula for the conditional state of the input field and the system $\mathcal{S}$ at time $\tau j$ that depends on all results of the measurement performed on the output field up to time $\tau j$. Note that, in general, the vectors $\ket{1_{\xi}}_{[j}$ and $\ket{1_{\phi}}_{[j}$ are not orthogonal. It is worth noting that the vector (\ref{condTH1}) has a simple and intuitive physical interpretation. Namely, we are dealing here with three possible scenarios. In the first scenario, the system has already encountered both photons, and it will interact in the future with the field being in the vacuum state.  In the second scenario, one of the photons has already appeared while the other one is still hidden in the input field. Finally, in the third scenario, all harmonic oscillators up to time $\tau j$ were in the ground state, and both photons will appear after time $\tau j$. In (\ref{condTH1}) we have a quantum superposition of terms corresponding to these scenarios. Moreover, it can be seen from (\ref{condTH1}) that, due to the interaction with the light, the system $\mathcal{S}$ becomes entangled with the input field. Taking the partial trace 
$\mathrm{Tr}_{\mathcal{E}_{[j}} \left(\ketbra{\tilde{\Psi}_{j}}{\tilde{\Psi}_{j}}\right)$, we obtain the reduced state of the system $\mathcal{S}$ at the time $j\tau$, 
\begin{equation}\label{condS}
\tilde{\rho}_{j}
=\frac{\rho_{j}}{\mathrm{Tr}_{\mathcal{S}}\rho_{j}},
\end{equation}
where
\begin{align}\label{condst3}
\rho_{j}&= \ketbra{\psi_{j}(0)}{\psi_{j}(0)}+\ketbra{\psi_{j}(1_{\xi})}{\psi_{j}(1_{\xi})}\sum_{k=j}^{M-1}\tau|\xi_{k}|^2\nonumber\\
&+\ketbra{\psi_{j}(1_{\xi})}{\psi_{j}(1_{\phi})}\sum_{k=j}^{M-1}\tau\xi_{k}\phi_{k}^{\ast}+\ketbra{\psi_{j}(1_{\phi})}{\psi_{j}(1_{\xi})}\sum_{k=j}^{M-1}\tau\phi_{k}\xi_{k}^{\ast}\nonumber\\
&+\ketbra{\psi_{j}(1_{\phi})}{\psi_{j}(1_{\phi})}\sum_{k=j}^{M-1}\tau|\phi_{k}|^2+\ketbra{\psi_{j}(2_{\xi\phi})}{\psi_{j}(2_{\xi\phi})}\frac{1}{\mathcal{N}_{\xi\phi}}\left|\sum_{k=j}^{M-1}\tau\xi_{k}\phi_{k}^{\ast}\right|^2\nonumber\\
&+\ketbra{\psi_{j}(2_{\xi\phi})}{\psi_{j}(2_{\xi\phi})}\frac{1}{\mathcal{N}_{\xi\phi}}\sum_{k=j}^{M-1}\tau|\phi_{k}|^2\sum_{k=j}^{M-1}\tau|\xi_{k}|^2.
\end{align}
The operator $\tilde{\rho}_{j}$ is the {\it a posteriori} state of $\mathcal{S}$ that depends on all results of the measurement up to $j\tau$ while the quantity $\mathrm{Tr}_{\mathcal{S}}{\rho}_{j}$ specifies the probability of particular trajectory. 

The conditional probability of detecting $n$ photons at the moment $(j+1)\tau$ given $\pmb{\eta}_j$ is defined by
\begin{equation}
p_{j+1}\left(n \Big|\pmb{\eta}_j \right)=\frac{\bra{\Psi_{j}} \hat{\mathbb{V}}_{[j}^{\dagger}\left(|n\rangle_{j}{}_{j}\langle n|\otimes \hat{\mathbbm{1}}_\mathcal{E}^{[j+1}\otimes \hat{\mathbbm{1}}_\mathcal{S}\right) \hat{\mathbb{V}}_{[j} \ket{\Psi_{j}}}{\braket{\Psi_{j}}{\Psi_{j}}}.
\end{equation}
From the form of $\ket{\Psi}_{j}$ and the property $\hat{V}_{ni}=\hat{O}(\sqrt{\tau}^{|n-i|})$, it follows that 
\begin{equation}
p_{j+1}\left(0\Big|\,\pmb{\eta}_j\right)=1-O(\tau),
\end{equation}
and for $n>1$ 
\begin{equation}
p_{j+1}\left(n\Big|\, \pmb{\eta}_j\right)=O(\tau^n).
\end{equation}
 We use here the Bachmann-Landau notation. We can conclude that the conditional probability of the lack of a photon in the time interval of length $\tau$ is the expression of the form $1-O(\tau)$ while the probability of registering one photon is $O(\tau)$. The conditional probability of detecting two or more photons is the term $O(\tau^2)$ and it goes to zero in the continuous-time limit. Thus, the processes of detection of two or more photons within the time interval of length $\tau$ are ignored. Therefore, we consider the realisation of the stochastic vector $\pmb{\eta}_j$ consisting only of the zeros and ones. By neglecting all terms of order more than one in $\tau$ and the terms associated with the processes of probability of order ${O}(\tau^2)$ or more, we ultimately obtain the set of four coupled stochastic recurrence equations
\begin{align}\label{TH-1_rec3}
	&\ket{\psi_{j+1}(0)} = \hat{V}_{\eta_{j+1}0}\ket{\psi_{j}(0)}+\sqrt{\tau}\xi_{j}\hat{V}_{\eta_{j+1}1}\ket{\psi_{j}(1_{\xi})}+\sqrt{\tau}\phi_{j}\hat{V}_{\eta_{j+1}1}\ket{\psi_{j}(1_{\phi})},\\
	&\ket{\psi_{j+1}(1_{\xi})} = \hat{V}_{\eta_{j+1}0}\ket{\psi_{j}(1_{\xi})}+\frac{1}{\sqrt{\mathcal{N}_{\xi\phi}}}\sqrt{\tau}\phi_{j}\hat{V}_{\eta_{j+1}1}\ket{\psi_{j}(2_{\xi\phi})},\\
	&\ket{\psi_{j+1}(1_{\phi})} =\hat{V}_{\eta_{j+1}0}\ket{\psi_{j}(1_{\phi})}+\frac{1}{\sqrt{\mathcal{N}_{\xi\phi}}}\sqrt{\tau}\xi_{j}\hat{V}_{\eta_{j+1}1}\ket{\psi_{j}(2_{\xi\phi})}
	,\\
	&\ket{\psi_{j+1}(2_{\xi\phi})} = \hat{V}_{\eta_{j+1}0}\ket{\psi_{j}(2_{\xi\phi})},\label{TH-1_rec4}
	\end{align}
where the system operators have the form
	\begin{align}\label{vmatrix}
	\hat{V}_{00}&= \hat{\mathbbm{1}}_{\mathcal{S}} - i\tau \hat{H}_{\mathcal{S}} - \tau \frac{1}{2}\hat{L}^\dagger \hat{L} + \hat{O}(\tau^{2}) ,\;\; \hat{V}_{11}=\hat{\mathbbm{1}}_{\mathcal{S}} + \hat{O}(\tau),\nonumber\\
	\hat{V}_{10}&=\sqrt{\tau} \hat{L} + \hat{O}(\tau^{3/2}),\;\; \hat{V}_{01}=- \sqrt{\tau} \hat{L}^\dagger + \hat{O}(\tau^{3/2})
	\end{align}
 and the random variable $\eta_{j+1}$ has only two possible realisation: $0$ and $1$. We would like to point out that the two sources of the recorded photons can be recognized in Eqs. (\ref{TH-1_rec3})--(\ref{TH-1_rec4}).  They are either emitted by the system $\mathcal{S}$ or originate directly from the input field. Clearly, these photons are indistinguishable.

Theorem \ref{TH-1} can be generalized to the case of the environment prepared in an arbitrary superposition of the two-photon vectors of the form (\ref{2phot}) or even more, for an arbitrary two-photon state vector in $\mathcal{H}_{\mathcal{E}}$. For details, refer to \ref{B}.  

\section{Quantum trajectories and photon statistics for unidirectional field}

In this section, we shall derive the formulae for quantum trajectories associated with the photon counting process. To this end we shall solve the set of stochastic recurrent equations (\ref{TH-1_rec3})--(\ref{TH-1_rec4}) and then determine their limits for the continuous time. Finally, we shall demonstrate how to determine the statistics of photon counts in the output field using quantum trajectories.

Let us notice that in order to describe the results of the photon counts, it is sufficient to indicate the moments at which these counts occurred. The string $(l_s,\ldots,l)$, where $0<l<\ldots <l_s\leq j$, means that exactly $s$ photons were detected at the moments $\tau_k = l_k\tau $ $(k=1,\ldots,s)$ and no other photons from time $0$ to $j\tau$. With the photon counting, we can associate the discrete stochastic process:
\begin{equation}\label{stoproc}
N_{j}=\sum_{k=1}^{j}\eta_{k},
\end{equation}
with the increment $\Delta N_{j}=N_{j+1}-N_{j}$. 
In order to present the formulae for quantum trajectories in a compact way, we introduce the following system operators: 
\begin{equation}
\hat{A}_{k}(\xi)=\hat{V}_{00}^{-k-1}\sqrt{\tau}\xi_{k}\hat{V}_{01}\hat{V}_{00}^{k},
\end{equation}
\begin{equation}
\hat{A}_{k}(\phi)=\hat{V}_{00}^{-k-1}\sqrt{\tau}\phi_{k}\hat{V}_{01}\hat{V}_{00}^{k},
\end{equation}
\begin{equation}
\hat{D}_{l}(\xi)=\hat{V}_{00}^{-l}\sqrt{\tau}\xi_{l-1}\hat{V}_{11}\hat{V}_{00}^{l-1},
\end{equation}
\begin{equation}
\hat{D}_{l}(\phi)=\hat{V}_{00}^{-l}\sqrt{\tau}\phi_{l-1}\hat{V}_{11}\hat{V}_{00}^{l-1},
\end{equation}
\begin{equation}
\hat{E}_{l}= \hat{V}_{00}^{-l}\hat{V}_{10}\hat{V}_{00}^{l-1},
\end{equation}
where $k=0,1,2,\ldots$ and $l=1,2,\ldots$. Let us observe that the operators $\hat{A}_{k}(\xi)$ and $\hat{A}_{k}(\phi)$ are associated with the absorption of a photon by the system $\mathcal{S}$ at time $k\tau$, while $\hat{D}_{l}(\xi)$ and $\hat{D}_{l}(\phi)$ are the operators related to detection at time $l\tau$ of photon coming directly from the input field.  The operator $\hat{E}_{l}$ describes the emission of a photon by the system $\mathcal{S}$ at time $l\tau$.  
The form of solution to the set of equations (\ref{TH-1_rec3})--(\ref{TH-1_rec4}) depends on the photon counting history. Below, we present the explicit formulae for the conditional vectors corresponding to some exemplary realisations of the stochastic process $N_{j}$:
\begin{enumerate}
	\item for detecting no photons from $0$ to $j\tau$:
	\begin{equation}\label{sol1}
	 \hspace{0pt} \ket{\psi_{j\vert0}(0)}= \frac{1}{\sqrt{\mathcal{N}_{\xi\phi}}}\hat{V}_{00}^{j}\sum_{k_2=1}^{j-1}\sum_{k_1=0}^{k_2-1}\left(\hat{A}_{k_2}(\xi)\hat{A}_{k_1}(\phi)+\hat{A}_{k_2}(\xi)\hat{A}_{k_1}(\phi)\right)\ket{\psi_{0}},
    \end{equation}
    \begin{equation}
	 \hspace{0pt} \ket{\psi_{j\vert0}(1_{\xi})}= \frac{1}{\sqrt{\mathcal{N}_{\xi\phi}}}\hat{V}_{00}^{j}\sum_{k=0}^{j-1}\hat{A}_{k}(\phi)\ket{\psi_{0}},
    \end{equation}
	\begin{equation}
     \hspace{0pt} \ket{\psi_{j\vert0}(1_{\phi})}= \frac{1}{\sqrt{\mathcal{N}_{\xi\phi}}}\hat{V}_{00}^{j}\sum_{k=0}^{j-1}\hat{A}_{k}(\xi)\ket{\psi_{0}},
    \end{equation}
    \begin{equation}
     \hspace{0pt} \ket{\psi_{j\vert0}(2_{\xi\phi})}=\hat{V}_{00}^{j}\ket{\psi_{0}},
	\end{equation}
	\item for the detection of a photon at time $l\tau$ and no other photons from $0$ to $j\tau$:
	\begin{align}
	\ket{\psi_{j\vert l}(0)}=& \frac{\hat{V}_{00}^j}{\sqrt{\mathcal{N}_{\xi\phi}}}
	\bigg[\hat{E}_{l}\sum_{k_2=1}^{l-2}\sum_{k_1=0}^{k_2-1}\left(\hat{A}_{k_2}(\xi)\hat{A}_{k_1}(\phi)+\hat{A}_{k_2}(\xi)\hat{A}_{k_1}(\phi)\right)\nonumber\\
	&+\sum_{k=l}^{j-1}\hat{A}_{k}(\xi)\hat{E}_{l}\sum_{k=0}^{l-2}\hat{A}_{k}(\phi)+\sum_{k=l}^{j-1}\hat{A}_{k}(\phi)\hat{E}_{l}\sum_{k=0}^{l-2}\hat{A}_{k}(\xi)\nonumber\\
	&+\sum_{k_2=l+1}^{j-1}\sum_{k_1=l}^{k_2-1}\left(\hat{A}_{k_2}(\xi)\hat{A}_{k_1}(\phi)+\hat{A}_{k_2}(\xi)\hat{A}_{k_1}(\phi)\right)\hat{E}_{l}\nonumber\\
	&+ \hat{D}_{l}(\xi)\sum_{k=0}^{l-2}\hat{A}_{k}(\phi)+\hat{D}_{l}(\phi)\sum_{k=0}^{l-2}\hat{A}_{k}(\xi)\nonumber\\
	&+\sum_{k=l}^{j-1}\hat{A}_{k}(\phi)\hat{D}_{l}(\xi)+\sum_{k=l}^{j-1}\hat{A}_{k}(\xi)\hat{D}_{l}(\phi)\bigg]\ket{\psi_{0}},
 \end{align}
 \begin{equation}
	 \hspace{0pt} \ket{\psi_{j\vert l}(1_{\xi})}= \frac{\hat{V}_{00}^{j}}{\sqrt{\mathcal{N}_{\xi\phi}}}
	\bigg[\hat{D}_{l}(\phi)+\hat{E}_{l}\sum_{k=0}^{l-2}\hat{A}_{k}(\phi)+\sum_{k=l}^{j-1}\hat{A}_{k}(\phi)\hat{E}_{l}\bigg]\ket{\psi_{0}},
    \end{equation}
	\begin{equation}
     \hspace{0pt} \ket{\psi_{j\vert l}(1_{\phi})}= \frac{\hat{V}_{00}^{j}}{\sqrt{\mathcal{N}_{\xi\phi}}}
	\bigg[\hat{D}_{l}(\xi)+\hat{E}_{l}\sum_{k=0}^{l-2}\hat{A}_{k}(\xi)+\sum_{k=l}^{j-1}\hat{A}_{k}(\xi)\hat{E}_{l}\bigg]\ket{\psi_{0}},
    \end{equation}
    \begin{equation}
	 \hspace{0pt} \ket{\psi_{j\vert l}(2_{\xi\phi})}= \hat{V}_{00}^{j}\hat{E}_{l}\ket{\psi_{0}}.\label{sol2}
	\end{equation}
	\end{enumerate}
Providing general formulae for the conditional vectors is challenging; however, stating the general rules of their structure is straightforward. In all terms of the conditional vectors, the total number of photon absorptions and direct detections from the input field is associated with the vectors as follows:
\begin{equation}
n_{A}+n_{D}= \begin{cases}
2\; \mathrm{for}\; \ket{\psi_{j|cond}(0)}\\
1\; \mathrm{for}\; \ket{\psi_{j|cond}(1_{\xi})}\; \mathrm{and}\; \ket{\psi_{j|cond}(1_{\phi})} \\
0\; \mathrm{for}\; \ket{\psi_{j|cond}(2_{\xi\phi})}\end{cases}.
\end{equation}
Furthermore, in all expressions of the conditional vectors, the total number of emissions and direct detections of photons from the input field is equal to the number of photon counts. 

We now turn to a description of the continuous evolution over time. We describe the evolution of the composed system up to time $t=M\tau$ such that $t$ is fixed and $\tau= t/M$ and we take the limit $M\to \infty$, thereby $\tau\to 0$. In this way, we obtain from (\ref{stoproc}) the continuous counting process $N_{t}$.
Let us notice that all realisation of the counting stochastic process $N_{t}$ may be divided into disjoint sectors which contain trajectories with exactly $s$ detected photons in the nonoverlapping intervals $[t_{1},t_{1}+dt_{1})$, $[t_{2},t_{2}+dt_{2})$, $\ldots$, $[t_{s},t_{s}+dt_{s})$ such that $t_0<t_1 < t_{2}<\ldots < t_s$. We deal now with the photon profiles $\xi, \phi\in L^{2}([t_0,+\infty))$ which fulfill the normalization condition:
\begin{equation}
\int_{t_0}^{+\infty}ds|\xi({s})|^2 =\int_{t_0}^{+\infty}ds|\phi({s})|^2=1,
\end{equation}
where by $t_0$ we denoted the moment when the interaction between the systems starts. 
Let us now introduce the system operators 
\begin{equation}\label{}
\hat{T}_t = e^{-i\hat{G}t},
\end{equation}
where $\hat{G}=\hat{H}_{\mathcal{S}}-\frac{i}{2}\hat{L}^{\dagger}\hat{L}$, and
\begin{equation}
\hat{A}_{t}(\xi)=-\hat{T}_{-t}\xi({t})\hat{L}^{\dagger}\hat{T}_{t},
\end{equation}
\begin{equation}
\hat{A}_{t}(\phi)=-\hat{T}_{-t}\phi({t})\hat{L}^{\dagger}\hat{T}_{t},
\end{equation}
\begin{equation}
\hat{E}_t = \hat{T}_{-t}\hat{L}\hat{T}_{t}.
\end{equation}
The operators $\hat{A}_{t}(\xi)$ and $\hat{A}_{t}(\phi)$ correspond to the absorption of photons at time $t$, while $\hat{E}_t$ refers to the emission of a photon by the system $\mathcal{S}$ at time $t$. The operator $\hat{T}_t$ refers to the evolution of the system $\mathcal{S}$ between the processes of absorption, emission, or detection. One can check, based on (\ref{sol1})--(\ref{sol2}), that in the continuous-time limit, we obtain the following formulae for the conditional vectors. Namely, 
\begin{enumerate}
	\item for zero counts from $t_0$ to $t$: 
	\begin{equation}\label{sol3}
	 \hspace{0pt} \ket{\psi_{t\vert 0}(0)}\!= \!\frac{1}{\sqrt{\mathcal{N}_{\xi\phi}}}\hat{T}_{t}\int_{t_0}^{t}ds^{\prime}\int_{t_0}^{s^{\prime}}ds\left(\hat{A}_{s^{\prime}}(\xi)\hat{A}_{s}(\phi) +\hat{A}_{s^{\prime}}(\xi)\hat{A}_{s}(\phi)\right)\hat{T}_{-t_0}\ket{\psi_{0}},
     \end{equation}
\begin{equation}
    \hspace{0pt} \ket{\psi_{t\vert0}(1_{\xi})}= \frac{1}{\sqrt{\mathcal{N}_{\xi\phi}}}\hat{T}_{t}\int_{t_0}^{t}ds\hat{A}_{s}(\phi)\hat{T}_{-t_0}\ket{\psi_{0}},
    \end{equation}
    \begin{equation}
 \hspace{0pt}	\ket{\psi_{t\vert0}(1_{\phi})}= \frac{1}{\sqrt{\mathcal{N}_{\xi\phi}}}\hat{T}_{t}\int_{t_0}^{t}ds\hat{A}_{s}(\xi)\hat{T}_{-t_0}\ket{\psi_{0}},
    \end{equation}
    \begin{equation}
 \hspace{0pt} \ket{\psi_{t\vert0}(2_{\xi\phi})}= \hat{T}_{t-t_{0}}\ket{\psi_{0}},
	\end{equation}
	
\item for a detection of one photon at time $t_1$ and no other photons from $t_0$ to $t$:
\begin{align}
\ket{\psi_{t\vert t_{1}}(0)}=& \frac{1}{\sqrt{\mathcal{N}_{\xi\phi}}}
\hat{T}_{t}\bigg[\hat{E}_{t_1}\int_{t_0}^{t_1}ds^{\prime}\int_{t_0}^{s^{\prime}}ds\left(\hat{A}_{s^{\prime}}(\xi)\hat{A}_{s}(\phi)+\hat{A}_{s^{\prime}}(\xi)\hat{A}_{s}(\phi)\right)\nonumber\\
&+\int_{t_1}^{t}
ds\hat{A}_{s}(\xi)\hat{E}_{t_1}
\int_{t_0}^{t_1}ds\hat{A}_{s}(\phi)\nonumber\\
&+\int_{t_1}^{t}ds\hat{A}_{s}(\phi)
\hat{E}_{l}\int_{t_0}^{t_1}ds\hat{A}_{s}(\xi)\nonumber\\
&+\int_{t_1}^{t}ds^{\prime}\int_{t_1}^{s^{\prime}}ds\left(\hat{A}_{s^{\prime}}(\xi)\hat{A}_{s}(\phi)+\hat{A}_{s^{\prime}}(\xi)\hat{A}_{s}(\phi)\right)\hat{E}_{t_1}\nonumber\\
&+ \xi({t_1})\int_{t_0}^{t}ds\hat{A}_{s}(\phi)+\phi({t_1})\int_{t_0}^{t}ds\hat{A}_{s}(\xi)\bigg]\hat{T}_{-t_0}\ket{\psi_{0}}
\end{align}
\begin{equation}
 \hspace{0pt} \ket{\psi_{t\vert t_1}(1_{\xi})}\!=\! \frac{1}{\sqrt{\mathcal{N}_{\xi\phi}}}
\hat{T}_{t}\bigg[\phi({t_1})+\hat{E}_{t_1}\int_{t_0}^{t_1}ds\hat{A}_{s}(\phi)+\int_{t_1}^{t}ds\hat{A}_{s}(\phi)\hat{E}_{t_1}\bigg]\hat{T}_{-t_0}\ket{\psi_{0}},
\end{equation}
\begin{equation}
 \hspace{0pt} \ket{\psi_{t\vert t_1}(1_{\phi})}= \frac{1}{\sqrt{\mathcal{N}_{\xi\phi}}}\hat{T}_{t}
\bigg[\xi({t_1})+\hat{E}_{t_1}\int_{t_0}^{t_1}ds\hat{A}_{s}(\xi)+\int_{t_1}^{t}ds\hat{A}_{s}(\xi)\hat{E}_{t_1}\bigg]\hat{T}_{-t_0}\ket{\psi_{0}},
\end{equation}
\begin{equation}
 \hspace{0pt} \ket{\psi_{t\vert t_1}(2_{\xi\phi})}= \hat{T}_{t}\hat{E}_{t_1}\hat{T}_{-t_0}\ket{\psi_{0}},
\end{equation}\label{sol4}
\end{enumerate}
where
\begin{equation}\label{factor}
 \mathcal{N}_{\xi\phi}=1+\left\vert\int_{t_0}^{+\infty}ds\xi^{\ast}({s})\phi(s)\right\vert^2.
\end{equation}

Let us observe that the conditional vectors allow us to determine both conditional and unconditional evolution of the system $\mathcal{S}$. Taking the average over all possible realisations of the counting process, we get the {\it a priori} evolution of the system $\mathcal{S}$. Thus the  {\it a priori} state of $\mathcal{S}$ can be written as 
\begin{equation}\label{apriori}
\sigma_{t}=\rho_{t\vert 0}+\sum_{s=1}^{+\infty}\int_{t_0}^{t}dt_{s}\int_{t_0}^{t_{s}}dt_{s-1}\ldots
\int_{t_0}^{t_{2}}dt_{1}\rho_{t\vert t_{s},t_{s-1},\ldots,t_{2},t_{1}},
\end{equation}
where each of the conditional operators depends on the registered trajectory from time $t_0$ to $t$ and has the following structure
\begin{align}
\rho_{t\vert cond}=&\ketbra{\psi_{t|cond}(0)}{\psi_{t|cond}(0)}+
\ketbra{\psi_{t|cond}(1_{\xi})}{\psi_{t|cond}(1_{\xi})}\int_{t}^{+\infty}ds|\xi({s})|^2\nonumber\\
&+\ketbra{\psi_{t|cond}(1_{\phi})}{\psi_{t|cond}(1_{\phi})}\int_{t}^{+\infty}ds|\phi({s})|^2\nonumber\\
&+\ketbra{\psi_{t|cond}(1_{\xi})}{\psi_{t|cond}(1_{\phi})}\int_{t}^{+\infty}ds\xi({s})\phi^{\ast}({s})+h.c. \nonumber\\
&+\ketbra{\psi_{t|cond}(2_{\xi\phi})}{\psi_{t|cond}(2_{\xi\phi})}\frac{1}{\mathcal{N}_{\xi\phi}}\left(\left|\int_{t}^{+\infty}ds\xi^{\ast}({s})\phi({s})\right|^2\right.\nonumber\\
&\left.+\int_{t}^{+\infty}ds|\phi({s})|^2\int_{t}^{+\infty}ds|\xi({s})|^2\right). 
\end{align}
We have found a decomposition the {\it a priori} state of $\mathcal{S}$ with respect to the counting stochastic process $N_{t}$. Let us emphasise that the formula (\ref{apriori}) for the reduced state of $\mathcal{S}$ is not unique, i.e., using homodyne or heterodyne measurement schemes, we would receive different stochastic representations for the reduced state of open system. 

The conditional vectors can be employed to determine the statistics of photon counting for the output field. The probability of not detecting any photons from time $t_0$ up to time $t$ is given by the formula
\begin{equation}\label{prob}
P_{t|0}=\mathrm{Tr}_{\mathcal{S}}\rho_{t|0},
\end{equation}
where $\rho_{t|0}$ is the conditional operator associated with the scenario of no photon counts from $t_0$ until $t$.
The exclusive probability density  \cite{ BarBel91,Srinivas81}, $ p_{t}(t_{s}, t_{s-1}, \ldots, t_{2}, t_{1})$, for exactly $s$ detections occurring from time $t_0$ to $t$ in the time intervals $[t_{1},t_{1}+dt_{1})$, $[t_{2},t_{2}+dt_{2})$, $\ldots$, $[t_{s},t_{s}+dt_{s})$, where $t_0<t_1 < t_{2}<\ldots < t_s<t$, is defined by  
\begin{equation}\label{probden}
p_{t}(t_{s}, t_{s-1},\ldots, t_{2}, t_{1})=\mathrm{Tr}_{\mathcal{S}}\rho_{t|t_{s}, t_{s-1},\ldots, t_{2}, t_{1}}.
\end{equation}
Note that from the quantities (\ref{prob}) and (\ref{probden}) one can obtain the whole statistics of the counting process. 
For instance,  the probability of detecting exactly $s$ photons from $t_0$ up to time $t$ is given by
\begin{equation}
P_{t}(s)=\int_{t_0}^{t}dt_{s}\int_{t_0}^{t_{s}}dt_{s-1}\ldots\int_{t_0}^{t_{2}}dt_{1}
p_{t}(t_{s}, t_{s-1}, \ldots  t_{2}, t_{1}) .
\end{equation}

\section{Repeated interactions model for bidirectional field}\label{BF}

In this section, we present the main properties of the collision model for a quantum system $\mathcal{S}$ interacting with a bidirectional electromagnetic field.  We describe the bidirectional field by the two chains of harmonic oscillators. Thus the Hilbert space of the environment is given by
\begin{equation}
\mathcal{H}_{\mathcal{E}}=\mathcal{H}_{\mathcal{E}_{1}}\otimes\mathcal{H}_{\mathcal{E}_{2}},
\end{equation}
where
\begin{equation}
\mathcal{H}_{\mathcal{E}_{l}}=\bigotimes_{k=0}^{M-1}\mathcal{H}_{\mathcal{E}_{l,k}},
\end{equation}
and $\mathcal{H}_{\mathcal{E}_{l,k}}$ is the Hilbert space associated with the harmonic oscillator of chain $l$ ($l=1,2$) that interacts with $\mathcal{S}$ in the interval $[k\tau, (k+1)\tau)$. One can imagine that the first chain describes the field going to the right, and the second refers to the field going to the left. We assume that the harmonic oscillators do not interact with each other but elements of each chain subsequently interact with the system $\mathcal{S}$ such that at a given moment $\mathcal{S}$ interacts with exactly two harmonic oscillators: one coming from the left and one coming from the right. The discrete evolution of the system composed of two chains of harmonic oscillators and the system 
$\mathcal{S}$ is described similarly to (\ref{disc_evol}), but the Hamiltonian containing the interaction of the system $\mathcal{S}$ with the field in the interval $[k\tau, (k+1)\tau)$ is given by
\begin{equation}
\hat{H}_{k} = \hat{H}_{\mathcal{S}}+\sum_{l=1}^{2}\frac{i}{\sqrt{\tau}}\left(\hat{b}_{l,k}^{\dagger}\otimes \hat{L}_{l}-\hat{b}_{l,k}\otimes \hat{L}^{\dagger}_{l}\right),
\end{equation}
where $\hat{L}_{l}\in \mathcal{B}(H_{S})$ for $l=1,2$. To simplify the notation, we omitted above the identity operators. 

We assume that the field is prepared in a two-photon separable state and the initial state of the composed system is given by the product state vector of the form
\begin{equation}\label{ini2}
\ket{\Psi_{0}}=\ket{1_{\xi}}\otimes\ket{1_{\phi}}\otimes\ket{\psi_{0}},
\end{equation}
where $\ket{1_{\xi}}$ and  $\ket{1_{\phi}}$ are the single-photon states associated with the fields labeled as first and second, respectively, and $\ket{\psi_{0}}$ is the initial state of $\mathcal{S}$.

\begin{remark}
An arbitrary two-photon state vector in $\mathcal{H}_{\mathcal{E}_{1}}\otimes\mathcal{H}_{\mathcal{E}_{2}}$ with photons located in different spatial modes can be written as 
\begin{equation}\label{arbitrary2}
\ket{2_{\Phi}}=\frac{1}{\sqrt{\mathcal{N}_{\Phi}}}\sum_{k_{1},k_{2}=0}^{M-1}
\tau \Phi_{k_{2}, k_{1}}\tilde{\hat{b}}_{2,k_{2}}^{\dagger}\tilde{\hat{b}}_{1,k_{1}}^{\dagger}\ket{vac}\otimes\ket{vac},\;\;\;l=1,2,
\end{equation}
where
\begin{equation}
\tilde{\hat{b}}_{l,k}^{\dagger}=\mathbbm{\hat{1}}_{\mathcal{E}_{l}}^{k-1]} \otimes \hat{b}_{l,k}^{\dagger}\otimes\mathbbm{\hat{1}}_{\mathcal{E}_{l}}^{[k+1},
\end{equation}
and $\mathcal{N}_{\Phi}=\sum_{k_1,k_2=0}^{M-1}\tau^2|\Phi_{k_2,k_1}|^2$ stands for the normalization factor. 
\end{remark}

\section{Repeated measurements and conditional state for bidirectional field}

In this section, we shall describe the stochastic evolution conditioned by the photon counting measurement of the two-dimensional output field. We consider measurements of the observables:
\begin{equation}\label{obs2}
\hat{b}_{l,k}^{\dagger}\hat{b}_{l,k},\;\;l=1,2;\; k=0, 1, \ldots M-1.
\end{equation}
In order to study the photon counting process, it is convenient to use for the unitary evolution operator the photon number representation,
\begin{equation}\label{operatorV}
\exp\left(-i\tau \hat{H}_{k}\right)  = \sum_{n_{1},n_{2},n_{3},n_{4}} |n_{1}n_{2}\rangle_{k} {}_{k}\langle n_{3}n_{4}| \otimes \hat{V}_{n_{1}n_{2},n_{3}n_{4}},
\end{equation}
where $\ket{n_{i_{1}}n_{i_{2}}}_{k}=\ket{n_{i_{1}}}_{k}\otimes \ket{n_{i_{2}}}_{k}$ are the number vectors in  $\mathcal{H}_{\mathcal{E}_{1,k}}\otimes\mathcal{H}_{\mathcal{E}_{2,k}}$. Here
$\hat{V}_{n_{1}n_{2},n_{3}n_{4}}\in \mathcal{B}(\mathcal{H_{S}})$ and their explicit forms are given in \ref{Inter_Oper}. 
The results of measurements performed on the output fields up to time $j\tau$ are now represented by the stochastic vector $\pmb{\eta}_j=(\eta_j,\eta_{j-1},\ldots,\eta_1)$, where $\eta_{k}=\left(\eta_{1,k},\eta_{2,k}\right)$ is the two-dimensional random variable with $\eta_{1,k}, \eta_{2,k}$ associated with the results of (\ref{obs2}).  
We formulate our observation on the conditional state of the input field and the system $\mathcal{S}$ in the following theorem.

\begin{Theorem}\label{TH-2}  The conditional state of the input parts of the environment (the part of the environment which has not interacted with $\mathcal{S}$ up to $j\tau$) and the system $\mathcal{S}$, for the initial state (\ref{ini2}) and the measurement of (\ref{obs2}), is at time $j\tau $ is given by 
	\begin{equation}\label{TH-2cond3}
	\ket{\tilde{\Psi}_{j}} = \frac{\ket{\Psi_{j}}}{\sqrt{\braket{\Psi_{j}}{\Psi_{j}}}},
	\end{equation}
	where $ \ket{\Psi_{j}}$ is the unnormalized conditional vector from the Hilbert space $\mathcal{H}_{\mathcal{E}}^{[j} \otimes \mathcal{H}_\mathcal{S}$ having the form
	\begin{align}\label{TH-2cond4}
	\ket{\Psi_{j}} =&\ket{vac}_{[j}\otimes \ket{vac}_{[j}\otimes \ket{\psi_{j}(00)}+ \ket{1_{\xi}}_{[j}\otimes\ket{vac}_{[j}\otimes \ket{\psi_{j}(1_{\xi}0)} \nonumber\\ 
&+ \ket{vac}_{[j}\otimes \ket{1_{\phi}}_{[j}\otimes\ket{\psi_{j}(01_{\phi})}
+	\ket{1_{\xi}}_{[j}\otimes \ket{1_{\phi}}_{[j}\otimes\ket{\psi_{j}(1_{\xi}1_{\phi})},
\end{align}
where $\ket{\psi_{j}(00)}$, $\ket{\psi_{j}(1_{\xi}0)}$, $\ket{\psi_{j}(01_{\phi})}$, and $\ket{\psi_{j}(1_{\xi}1_{\phi})}$ are the conditional vectors from the Hilbert space $\mathcal{H}_{\mathcal{S}}$ which satisfy the set of coupled recurrence equations:
		\begin{align}\label{TH-2rec2}
	\ket{\psi_{j+1}(00)} = &\hat{V}_{\eta_{j+1},00}\ket{\psi_{j}(00)}+\sqrt{\tau}\xi_{j}\hat{V}_{\eta_{j+1},10}\ket{\psi_{j}(1_{\xi}0)}
	\nonumber\\&+\sqrt{\tau}\phi_{j}\hat{V}_{\eta_{j+1},01}\ket{\psi_{j}(01_{\phi})}+\tau\xi_{j}\phi_{j}\hat{V}_{\eta_{j+1},11}\ket{\psi_{j}(1_{\xi}1_{\phi})},\\
	\ket{\psi_{j+1}(1_{\xi}0)} =& \hat{V}_{\eta_{j+1},00}\ket{\psi_{j}(1_{\xi}0)}+\sqrt{\tau}\phi_{j}\hat{V}_{\eta_{j+1},01}\ket{\psi_{j}(1_{\xi}1_{\phi})},\\
	\ket{\psi_{j+1}(01_{\phi})} =&\hat{V}_{\eta_{j+1},00}\ket{\psi_{j}(01_{\phi})}+\sqrt{\tau}\xi_{j}\hat{V}_{\eta_{j+1},10}\ket{\psi_{j}(1_{\xi}1_{\phi})},
	\\
	\ket{\psi_{j+1}(1_{\xi}1_{\phi})} =&\hat{V}_{\eta_{j+1},00}\ket{\psi_{j}(1_{\xi}1_{\phi})}\label{TH-2rec3}
	\end{align}
and initially $\ket{\psi_{j=0}(00)}=\ket{\psi_{j=0}(1_{\xi}0)}=\ket{\psi_{j=0}(01_{\phi})}=0$, $\ket{\psi_{j=0}(1_{\xi}1_{\phi})}=\ket{\psi_{0}}$. 
  \end{Theorem}
For the proof of Theorem \ref{TH-2}, see \ref{ProofTH-2}. The interpretation of the above conditional state is similar to that given for the vector (\ref{condTH1}) in the previous section, for the unidirectional field, but note that here all vectors of the input field, which can be briefly written down as $\ket{vac, vac}_{[j}$, $\ket{1_{\xi}, vac}_{[j}$, $\ket{vac, 1_{\phi}}_{[j}$, $\ket{1_{\xi}, 1_{\phi}}_{[j}$, are mutually orthogonal. 
The system $\mathcal{S}$, as before, becomes entangled with the input field, and for this reason, its evolution becomes non-Markovian. Eliminating from the description the degrees of freedom associated with the input field, we obtain the {\it a posteriori} state of $\mathcal{S}$, $\tilde{\rho}_{j}=\rho_{j}/\mathrm{Tr}_{\mathcal{S}}\rho_{j}$, where $\rho_{j}$ is the conditional operator having the form
\begin{align}
\rho_{j}=& \ketbra{\psi_{j}(00)}{\psi_{j}(00)}+\ketbra{\psi_{j}(1_{\xi}0)}{\psi_{j}(1_{\xi}0)}\sum_{k=j}^{M-1}\tau|\xi_{k}|^2+\ketbra{\psi_{j}(01_{\phi})}{\psi_{j}(01_{\phi})}\sum_{k=j}^{M-1}\tau|\phi_{k}|^2\nonumber\\
&+\ketbra{\psi_{j}(1_{\xi}1_{\phi})}{\psi_{j}(1_{\xi}1_{\phi})}\sum_{k=j}^{M-1}\tau|\phi_{k}|^2\sum_{k=j}^{M-1}\tau|\xi_{k}|^2.
\end{align}

Let us introduce now the two-dimensional discrete stochastic process
\begin{equation}\label{sp}
N_{j}=\left(N_{1,j},N_{2,j}\right)
\end{equation}
where 
\begin{equation}
N_{1,j}=\sum_{k=1}^{j}\eta_{1,j}, \;\; N_{2,j}=\sum_{k=1}^{j}\eta_{2,j} 
\end{equation}
are the processes referring to the counts registered respectively by the right and the left detector. 
The conditional probability of the outcome $(n,n^{\prime})$ at $(j+1)\tau$ given the sequence $\pmb{\eta}_{j}$ is defined by  
\begin{equation}
p_{j+1}\left((n,n^{\prime})|\pmb{\eta}_j\right)=
\frac{ \bra{\Psi_{j+1}}\hat{\mathbb{V}}_{[j}^{\dagger}\left(|nn^{\prime}\rangle_{j}{}_{j}\langle nn^{\prime}|\otimes \hat{\mathbbm{1}}_\mathcal{E}^{[j+1}\otimes \hat{\mathbbm{1}}_\mathcal{S}\right) \hat{\mathbb{V}}_{[j} \ket{\Psi_{j+1}} }{\braket{\Psi_{j}}{\Psi_{j}}}.
\end{equation}
By applying the result of Theorem \ref{TH-2} and the fact that $\hat{V}_{nn^{\prime},ii^{\prime}}=\hat{O}(\sqrt{\tau}^{|n-i|+|n^{\prime}-i^{\prime}|})$, one can check that:
\begin{equation}
p_{j+1}\left((0,0)|\pmb{\eta}_j\right)= 1-O(\tau),
\end{equation}
\begin{equation}
p_{j+1}\left((1,0)|\pmb{\eta}_j\right)=O(\tau),
\end{equation}
\begin{equation}
p_{j+1}\left((0,1)|\pmb{\eta}_j\right)=O(\tau),
\end{equation}
and in a general case for $n,n^{\prime}$ 
\begin{equation}
p_{j+1}\left((n,n^{\prime})|\pmb{\eta}_j\right)=O(\tau^{n+n^{\prime}}).
\end{equation} 
It follows that, most of the time, we observe two vacuums (zero counts on both detectors) and from time to time we detect a count on the left or on the right. The simultaneous counts in both detectors occur with the probability of $O(\tau^2)$. The probability of such detection is equal to zero in the continuous-time limit and we ignore this process. Thus the random variable $\eta_{j+1}$ has only three possible realisation: $(0,0)$, $(1,0)$, or $(0,1)$. Neglecting in Eqs. (\ref{TH-2rec2})--(\ref{TH-2rec3}) all terms of order more than one in $\tau$ and the terms associated with the processes of probability of ${O}(\tau^2)$ or more, we obtain the set of four coupled stochastic recurrence equations:
\begin{align}\label{TH-2rec4}
	\ket{\psi_{j+1}(00)} = &\hat{V}_{\eta_{j+1},00}\ket{\psi_{j}(00)}+\sqrt{\tau}\xi_{j}\hat{V}_{\eta_{j+1},10}\ket{\psi_{j}(1_{\xi}0)}
	+\sqrt{\tau}\phi_{j}\hat{V}_{\eta_{j+1},01}\ket{\psi_{j}(01_{\phi})},\\
	\ket{\psi_{j+1}(1_{\xi}0)} =& \hat{V}_{\eta_{j+1},00}\ket{\psi_{j}(1_{\xi}0)}+\sqrt{\tau}\phi_{j}\hat{V}_{\eta_{j+1},01}\ket{\psi_{j}(1_{\xi}1_{\phi})},\\
	\ket{\psi_{j+1}(01_{\phi})} =&\hat{V}_{\eta_{j+1},00}\ket{\psi_{j}(01_{\phi})}+\sqrt{\tau}\xi_{j}\hat{V}_{\eta_{j+1},10}\ket{\psi_{j}(1_{\xi}1_{\phi})},
	\\
	\ket{\psi_{j+1}(1_{\xi}1_{\phi})} =&\hat{V}_{\eta_{j+1},00}\ket{\psi_{j}(1_{\xi}1_{\phi})}\label{TH-2rec5}.
	\end{align}

In \ref{GeneralizationTH-2}, we provide a generalization of Theorem \ref{TH-2} to the case of a two-photon state vector in $\mathcal{H}_{\mathcal{E}_{1}}\otimes\mathcal{H}_{\mathcal{E}_{2}}$ defined by (\ref{arbitrary2}).

\section{Quantum trajectories and photon statistics for bidirectional field}

In this section, we will present exemplary solutions to the set of Eqs. (\ref{TH-2rec4})-(\ref{TH-2rec5}) and describe the structure of the arbitrary solutions. We will also describe formulas for conditional vectors for continuous-in-time evolution and field observations. 

Let us establish that we measure the first field from the right and the second field from the left. We shall use the notation $R$ and $L$ to denote the right and left detectors, respectively. In order to describe the realisation of the stochastic process (\ref{sp}), we need to indicate the moment of counting and the referring detector (either the left or the right).  Let us notice that each realisation of the stochastic process (\ref{sp}) can be written in the form
\begin{equation}
\left(D_{m},l_{m};\ldots;D_{2},l_{2};D_{1},l_{1}\right).
\end{equation}
which provides the following information: the first photon was recorded at time $l_{1}\tau$ by the detector $D_1$, the second photon at time $l_{2}\tau$ by the detector $D_2$ and so on, where  $l_1<l_2<\ldots<l_{m}$, and no other photons were detected from $0$ to $j\tau$. Let us introduce now the system operators:
\begin{equation}
\hat{A}_{k}(\xi)=\hat{V}_{00,00}^{-k-1}\sqrt{\tau}\xi_{k}\hat{V}_{00,10}\hat{V}_{00,00}^{k},
\end{equation}
\begin{equation}
\hat{A}_{k}(\phi)=\hat{V}_{00,00}^{-k-1}\sqrt{\tau}\phi_{k}\hat{V}_{00,01}\hat{V}_{00,00}^{k},
\end{equation}
\begin{equation}
\hat{D}_{l}(\xi)=\hat{V}_{00}^{-l}\sqrt{\tau}\xi_{l-1}\hat{V}_{10,10}\hat{V}_{00}^{l-1},
\end{equation}
\begin{equation}
\hat{D}_{l}(\phi)=\hat{V}_{00,00}^{-l}\sqrt{\tau}\phi_{l-1}\hat{V}_{01,01}\hat{V}_{00,00}^{l-1},
\end{equation}
\begin{equation}
\hat{E}_{l}(R)= \hat{V}_{00,00}^{-l}\hat{V}_{10,00}\hat{V}_{00,00}^{l-1},
\end{equation}
\begin{equation}
\hat{E}_{l}(L)= \hat{V}_{00,00}^{-l}\hat{V}_{01,00}\hat{V}_{00,00}^{l-1},
\end{equation}
where $k=0,1,2,\ldots$ and $l=1,2,\ldots$. Note that the operators $\hat{A}_{k}(\xi)$ and $\hat{A}_{k}(\phi)$ are related to the absorption of photons coming from the left and right, respectively. The operators $\hat{D}_{l}(\xi)$ and $\hat{D}_{l}(\phi)$ are related to the direct count of the photon coming from the left and right.  $\hat{E}_{l}(R)$ and $\hat{E}_{l}(L)$ describe the emission of a photon by the system $\mathcal{S}$ to the right and left. We present the conditional vectors for three simple trajectories:
\begin{enumerate}
	\item for detecting no photons from $0$ to $j\tau$:
	\begin{equation}\label{sol5}
	\ket{\psi_{j\vert0}(00)}= \hat{V}_{00,00}^{j}\sum_{k_2=1}^{j-1}\sum_{k_1=0}^{k_2-1}\left(\hat{A}_{k_2}(\xi)\hat{A}_{k_1}(\phi)+\hat{A}_{k_2}(\xi)\hat{A}_{k_1}(\phi)\right)\ket{\psi_{0}},
    \end{equation}
    \begin{equation}
	\ket{\psi_{j\vert0}(1_{\xi}0)}= \hat{V}_{00,00}^{j}\sum_{k=0}^{j-1}\hat{A}_{k}(\phi)\ket{\psi_{0}},
    \end{equation}
    \begin{equation}
	\ket{\psi_{j\vert0}(01_{\phi})}= \hat{V}_{00,00}^{j}\sum_{k=0}^{j-1}\hat{A}_{k}(\xi)\ket{\psi_{0}},
    \end{equation}
    \begin{equation}
	\ket{\psi_{j\vert0}(1_{\xi}1_{\phi})}=\hat{V}_{00,00}^{j}\ket{\psi_{0}},
	\end{equation}
	\item for the detection of a photon at time $l\tau$ on the right hand side and no other photons from $0$ to $j\tau$:
	\begin{align}
	\ket{\psi_{j\vert (R,l)}(0)}=& \hat{V}_{00,00}^{j}
	\bigg[\hat{E}_{l}(R)\sum_{k_2=1}^{l-2}\sum_{k_1=0}^{k_2-1}\left(\hat{A}_{k_2}(\xi)\hat{A}_{k_1}(\phi)+\hat{A}_{k_2}(\xi)\hat{A}_{k_1}(\phi)\right)\nonumber\\
	&+\sum_{k=l}^{j-1}\hat{A}_{k}(\xi)\hat{E}_{l}(R)\sum_{k=0}^{l-2}\hat{A}_{k}(\phi)+\sum_{k=l}^{j-1}\hat{A}_{k}(\phi)\hat{E}_{l}(R)\sum_{k=0}^{l-2}\hat{A}_{k}(\xi)\nonumber\\
	&+\sum_{k_2=l+1}^{j-1}\sum_{k_1=l}^{k_2-1}\left(\hat{A}_{k_2}(\xi)\hat{A}_{k_1}(\phi)+\hat{A}_{k_2}(\xi)\hat{A}_{k_1}(\phi)\right)\hat{E}_{l}(R)\nonumber\\
	&+ \hat{D}_{l}(\xi)\sum_{k=0}^{l-2}\hat{A}_{k}(\phi)+\sum_{k=l}^{j-1}\hat{A}_{k}(\phi)\hat{D}_{l}(\xi)\bigg]\ket{\psi_{0}},
 \end{align}
 \begin{equation}
	\ket{\psi_{j\vert (R,l)}(1_{\xi}0)}= \hat{V}_{00,00}^{j}
	\bigg[\hat{E}_{l}(R)\sum_{k=0}^{l-2}\hat{A}_{k}(\phi)+\sum_{k=l}^{j-1}\hat{A}_{k}(\phi)\hat{E}_{l}(R)\bigg]\ket{\psi_{0}},
    \end{equation}
    \begin{equation}
	\ket{\psi_{j\vert (R,l)}(01_{\phi})}= \hat{V}_{00,00}^{j}	\bigg[\hat{D}_{l}(\xi)+\hat{E}_{l}(R)\sum_{k=0}^{l-2}\hat{A}_{k}(\xi)+\sum_{k=l}^{j-1}\hat{A}_{k}(\xi)\hat{E}_{l}(R)\bigg]\ket{\psi_{0}},
    \end{equation}
    \begin{equation}
	\ket{\psi_{j\vert (R,l)}(1_{\xi}1_{\phi})}= \hat{V}_{00,00}^{j}\hat{E}_{l}(R)\ket{\psi_{0}},
	\end{equation}
 \item for the detection of a photon at time $l\tau$ on the left hand side and no other photons from $0$ to $j\tau$:
	\begin{align}
	\ket{\psi_{j\vert (L,l)}(0)}=&\hat{V}_{00,00}^j
	\bigg[\hat{E}_{l}(L)\sum_{k_2=1}^{l-2}\sum_{k_1=0}^{k_2-1}\left(\hat{A}_{k_2}(\xi)\hat{A}_{k_1}(\phi)+\hat{A}_{k_2}(\xi)\hat{A}_{k_1}(\phi)\right)\nonumber\\
	&+\sum_{k=l}^{j-1}\hat{A}_{k}(\xi)\hat{E}_{l}(L)\sum_{k=0}^{l-2}\hat{A}_{k}(\phi)+\sum_{k=l}^{j-1}\hat{A}_{k}(\phi)\hat{E}_{l}(L)\sum_{k=0}^{l-2}\hat{A}_{k}(\xi)\nonumber\\
	&+\sum_{k_2=l+1}^{j-1}\sum_{k_1=l}^{k_2-1}\left(\hat{A}_{k_2}(\xi)\hat{A}_{k_1}(\phi)+\hat{A}_{k_2}(\xi)\hat{A}_{k_1}(\phi)\right)\hat{E}_{l}(L)\nonumber\\
	&+\hat{D}_{l}(\phi)\sum_{k=0}^{l-2}\hat{A}_{k}(\xi)+\sum_{k=l}^{j-1}\hat{A}_{k}(\xi)\hat{D}_{l}(\phi)\bigg]\ket{\psi_{0}},
 \end{align}
 \begin{equation}
	\ket{\psi_{j\vert (L,l)}(1_{\xi}0)}= \hat{V}_{00,00}^{j}
	\bigg[\hat{D}_{l}(\phi)+\hat{E}_{l}(L)\sum_{k=0}^{l-2}\hat{A}_{k}(\phi)+\sum_{k=l}^{j-1}\hat{A}_{k}(\phi)\hat{E}_{l}(L)\bigg]\ket{\psi_{0}},
    \end{equation}
    \begin{equation}
	\ket{\psi_{j\vert (L,l)}(01_{\phi})}= \hat{V}_{00,00}^{j}
	\bigg[\hat{E}_{l}(L)\sum_{k=0}^{l-2}\hat{A}_{k}(\xi)+\sum_{k=l}^{j-1}\hat{A}_{k}(\xi)\hat{E}_{l}(L)\bigg]\ket{\psi_{0}},
    \end{equation}
    \begin{equation}
	\ket{\psi_{j\vert (L,l)}(1_{\xi}1_{\phi})}= \hat{V}_{00,00}^{j}\hat{E}_{l}(L)\ket{\psi_{0}}.
	\end{equation}
	\end{enumerate}
 Let us observe that for all expressions in the formulae for the conditional vectors, the total number of the direct photon detections from the input field running in a given direction and absorptions of photons by the system $\mathcal{S}$ satisfies the following conditions:
 \begin{equation}
n_{A}+n_{D}= \begin{cases}
2\; \mathrm{for}\; \ket{\psi_{j|cond}(00)}\\
1\; \mathrm{for}\; \ket{\psi_{j|cond}(1_{\xi}0)}\; \mathrm{and}\; \ket{\psi_{j|cond}(01_{\phi})} \\
0\; \mathrm{for}\; \ket{\psi_{j|cond}(1_{\xi}1_{\phi})}\end{cases}.
\end{equation}
Moreover, the total number of emissions in a given direction and direct detections from the input field running in a given direction is equal to the total number of counts on that side. 

In the continuous-time limit, we obtain from (\ref{sp}) the two-dimensional stochastic process 
\begin{equation}\label{csp}
N_t=\left(N_{1,t},N_{2,t}\right),
\end{equation}  
which describes the continuous in-time detection of photons in the right and left detectors. To characterize the stochastic counting process by the exclusive probability densities, we present the analytical formulae for the conditional vectors associated with different realisations of (\ref{csp}). For this purpose, we introduce the system operators 
\begin{equation}\label{}
\hat{T}_t = e^{-i\hat{G}t},
\end{equation}
where $\hat{G}=\hat{H}_{\mathcal{S}}-\frac{i}{2}(\hat{L}^{\dagger}_{1}\hat{L}_{1}+\hat{L}^{\dagger}_{2}\hat{L}_{2})$, and
\begin{equation}
\hat{A}_{t}(\xi)=-\hat{T}_{-t}\xi(t)\hat{L}_{1}^{\dagger}\hat{T}_{t},
\end{equation}
\begin{equation}
\hat{A}_{t}(\phi)=-\hat{T}_{-t}\phi(t)\hat{L}_{2}^{\dagger}\hat{T}_{t},
\end{equation}
\begin{equation}\label{}
\hat{E}_t(R) = \hat{T}_{-t}\hat{L}_{1}\hat{T}_{t}.
\end{equation}
\begin{equation}\label{}
\hat{E}_t(L) = \hat{T}_{-t}\hat{L}_{2}\hat{T}_{t}.
\end{equation}
Note that the operators $\hat{E}_t({R})$ and $\hat{E}_t({L})$ are related to the emission of photon by the system $\mathcal{S}$, respectively to the right and the left. We present below the conditional vector for chosen trajectories. In the continuous-time limit, we obtain the conditional vectors at time $t$ of the form:
\begin{enumerate}
	\item for zero photons from $0$ to $t$: 
	\begin{equation}
	\ket{\psi_{t\vert 0}(00)}=\hat{T}_{t}\int_{t_0}^{t}ds^{\prime}\int_{t_0}^{s^{\prime}}ds\left(\hat{A}_{s^{\prime}}(\xi)\hat{A}_{s}(\phi)+\hat{A}_{s^{\prime}}(\xi)\hat{A}_{s}(\phi)\right)\hat{T}_{-t_0}\ket{\psi_{0}},\label{convec}
    \end{equation}
    \begin{equation}
	\ket{\psi_{t\vert0}(1_{\xi}0)}=\hat{T}_{t}\int_{t_0}^{t}ds\hat{A}_{k}(\phi)\hat{T}_{-t_0}\ket{\psi_{0}},
    \end{equation}
    \begin{equation}
	\ket{\psi_{t\vert0}(01_{\phi})}=\hat{T}_{t}\int_{t_0}^{t}ds\hat{A}_{s}(\xi)\hat{T}_{-t_0}\ket{\psi_{0}},
    \end{equation}
    \begin{equation}
	\ket{\psi_{j\vert0}(1_{\xi}1_{\phi})}=\hat{T}_{t-t_0}\ket{\psi_{0}},
	\end{equation}
\item for the detection of a photon at time $t_1$ on the right hand side and no other photons from $0$ to $t$:
\begin{align}
\ket{\psi_{t\vert (R,t_{1})}(00)}=&
\hat{T}_{t}\bigg[\hat{E}_{t_1}(R)\int_{t_0}^{t_1}ds^{\prime}\int_{0}^{s^{\prime}}ds\left(\hat{A}_{s^{\prime}}(\xi)\hat{A}_{s}(\phi)+\hat{A}_{s^{\prime}}(\xi)\hat{A}_{s}(\phi)\right)\nonumber\\
&+\int_{t_1}^{t}
ds\hat{A}_{s}(\xi)\hat{E}_{t_1}(R)
\int_{t_0}^{t_1}ds\hat{A}_{s}(\phi)
\nonumber\\&+\int_{t_1}^{t}ds\hat{A}_{s}(\phi)
\hat{E}_{l}(R)\int_{t_0}^{t_1}ds\hat{A}_{s}(\xi)\nonumber\\
&+\int_{t_1}^{t}ds^{\prime}\int_{t_1}^{s^{\prime}}ds\left(\hat{A}_{s^{\prime}}(\xi)\hat{A}_{s}(\phi)+\hat{A}_{s^{\prime}}(\xi)\hat{A}_{s}(\phi)\right)\hat{E}_{t_1}(R)\nonumber\\
&+ \xi({t_1})\int_{t_0}^{t}ds\hat{A}_{s}(\phi)\bigg]\hat{T}_{-t_0}\ket{\psi_{0}},
\end{align}
\begin{equation}
\ket{\psi_{t\vert (R,t_1)}(1_{\xi}0)}= 
\hat{T}_{t}\bigg[\hat{E}_{t_1}(R)\int_{t_0}^{t_1}ds\hat{A}_{s}(\xi)+\int_{t_1}^{t}\hat{A}_{s}(\xi)\hat{E}_{t_1}(R)\bigg]\hat{T}_{-t_0}\ket{\psi_{0}},
\end{equation}
\begin{align}
\ket{\psi_{t\vert (R,t_1)}(01_{\phi})}= &\hat{T}_{t}
\bigg[\xi(t_1)+\hat{E}_{t_1}(R)\int_{t_0}^{t_1}ds\hat{A}_{s}(\phi)\nonumber\\&+\int_{t_1}^{t}ds\hat{A}_{s}(\phi)\hat{E}_{t_1}(R)\bigg]\hat{T}_{-t_0}\ket{\psi_{0}},
\end{align}
\begin{equation}
\ket{\psi_{j\vert (R,t_1)}(1_{\xi}1_{\phi})}=\hat{T}_{t}\hat{E}_{t_1}(R)\hat{T}_{-t_0}\ket{\psi_{0}},
\end{equation}	

\item for the detection of a photon at time $t_1$ on the left hand side and no other photons from $0$ to $t$:
\begin{align}\label{sol7}
\ket{\psi_{t\vert (L,t_{1})}(00)}=&
\hat{T}_{t}\bigg[\hat{E}_{t_1}(L)\int_{t_0}^{t_1}ds^{\prime}\int_{t_0}^{s^{\prime}}ds\left(\hat{A}_{s^{\prime}}(\xi)\hat{A}_{s}(\phi)+\hat{A}_{s^{\prime}}(\xi)\hat{A}_{s}(\phi)\right)\nonumber\\
&+\int_{t_1}^{t}
ds\hat{A}_{s}(\xi)\hat{E}_{t_1}(L)
\int_{t_0}^{t_1}ds\hat{A}_{s}(\phi)
\nonumber\\&+\int_{t_1}^{t}ds\hat{A}_{s}(\phi)
\hat{E}_{l}(L)\int_{t_0}^{t_1}ds\hat{A}_{s}(\xi)\nonumber\\
&+\int_{t_1}^{t}ds^{\prime}\int_{t_1}^{s^{\prime}}ds\left(\hat{A}_{s^{\prime}}(\xi)\hat{A}_{s}(\phi)+\hat{A}_{s^{\prime}}(\xi)\hat{A}_{s}(\phi)\right)\hat{E}_{t_1}(L)\nonumber\\
&+\phi({t_1})\int_{t_0}^{t}ds\hat{A}_{s}(\xi)\bigg]\hat{T}_{-t_0}\ket{\psi_{0}},
\end{align}
\begin{align}
\ket{\psi_{t\vert (L,t_1)}(1_{\xi})}=& 
\hat{T}_{t}\bigg[\xi(t_1)+\hat{E}_{t_1}(L)\int_{t_0}^{t_1}ds\hat{A}_{s}(\xi)\nonumber\\
&+\int_{t_1}^{t}\hat{A}_{s}(\xi)\hat{E}_{t_1}(L)\bigg]\hat{T}_{-t_0}\ket{\psi_{0}},
\end{align}
\begin{equation}
\ket{\psi_{t\vert (L,t_1)}(01_{\phi})}= \hat{T}_{t}
\bigg[\hat{E}_{t_1}(L)\int_{t_0}^{t_1}ds\hat{A}_{s}(\xi)+\int_{t_1}^{t}ds\hat{A}_{s}(\xi)\hat{E}_{t_1}(L)\bigg]\hat{T}_{-t_0}\ket{\psi_{0}},
\end{equation}
\begin{equation}
\ket{\psi_{j\vert (L,t_1)}(1_{\xi}1_{\phi})}=\hat{T}_{t}\hat{E}_{t_1}(L)\hat{T}_{-t_0}\ket{\psi_{0}}.
\end{equation}	
\end{enumerate}

The {\it a priori} state of $\mathcal{S}$ in the continuous-time limit can be written in the form 
\begin{equation}\label{apriori2}
\sigma_{t}=\rho_{t\vert 0}+\sum_{s=1}^{+\infty}\sum_{D_{s},\ldots, D_{2},D_{1}=R,L}\int_{t_0}^{t}dt_{s}\int_{t_0}^{t_{s}}dt_{s-1}\ldots
\int_{t_0}^{t_{2}}dt_{1}\rho_{j|D_{t_s},t_{s};\ldots;D_{2},t_{2};D_{1},t_{1}}
\end{equation}
with the conditional operators: 
\begin{align}
\rho_{t\vert cond}=&\ketbra{\psi_{t|cond}(00)}{\psi_{t|cond}(00)}+
\ketbra{\psi_{t|cond}(1_{\xi}0)}{\psi_{t|cond}(1_{\xi}0)}\int_{t}^{+\infty}ds|\xi({s})|^2\nonumber\\
&+\ketbra{\psi_{t|cond}(01_{\phi})}{\psi_{t|cond}(01_{\phi})}\int_{t}^{+\infty}ds|\phi({s})|^2\nonumber\\
&+\ketbra{\psi_{t|cond}(1_{\xi}1_{\phi})}{\psi_{t|cond}(1_{\xi}1_{\phi})}\int_{t}^{+\infty}ds|\phi({s})|^2\int_{t}^{+\infty}ds|\xi({s})|^2. 
\end{align}
Note that the sum in formula (\ref{apriori2}) is taken over all the pathways of photon detection occurring in the interval from $t_0$ to $t$. The exclusive probability density of $s$ counts occurring in the nonoverlapping intervals $[t_{1}+dt_1), [t_{2}+dt_2),\ldots, [t_{s}+dt_{s})$, such that  $t_0<t_{1}< t_{2}<\ldots< t_{s}$, taking place respectively in the detectors  $D_{1}, D_{2}, \ldots, D_{s}$, and no other detections in the interval from $t_0$ to $t$ is defined by  
\begin{eqnarray}\label{condensity}
p_{t|{D_{m}},t_{m};\ldots;{D_{2}},t_{2};{D_1},t_{1}}=
\mathrm{Tr}_{\mathcal{S}}\rho_{t|{D_{s}},t_{s};\ldots;D_{2},t_{2};{D_1},t_{1}}
\end{eqnarray}
Thus the probability of $s$ counts registered in the interval from $t_0$ to $t$ at the detectors $D_{1}, D_2, \ldots, D_{s}$ is given by 
	\begin{equation}
{P_{t|s;D_{s},\ldots,D_2,D_{1}}=}\int_{0}^{t}dt_{s}\int_{0}^{t_{s}}dt_{s-1}\ldots
	\int_{0}^{t_{2}}dt_{1}
	 p_{t|{D_{s}},t_{s};\ldots;{D_{2}},t_{2};{D_1},t_{1}}.
	\end{equation}
Let us emphasize that the exclusive probability density enables us to find the whole statistics of the photon counts in the output field.

\section{Probability of two-photon absorption for a three-level atom in a ladder configuration}

As an example of a system interacting with a propagating field in a two-photon state, we consider a three-level atom in a ladder configuration. We denote the states of the atom by $\ket{g}$, $\ket{e}$, $\ket{f}$. 
Our objective is to show how the probability of two-photon absorption can be determined by means of the conditional vectors defining the quantum trajectories. We assume that the atom is initially in the ground state and derive a formula for the probability that the system will be in the excited state $\ket{f}$ at a given moment. We analyze two scenarios: the first, for the atom excited by a unidirectional two-photon field, and the second, involving a bidirectional two-photon field. Furthermore, we will derive formulas for two-photon states that optimally excite a three-level system of a ladder-type level structure, i.e., we will determine the two-photon states that maximize the absorption probability at a given moment.

\subsection{Results for unidirectional field in two-photons state}

Let us start with the case when the unidirectional field is prepared in the two-photon state of the form 
\begin{equation}\label{tps1}
\ket{2_{\xi\phi}} =\frac{1}{\sqrt{\mathcal{N}_{\xi\phi}}}\int_{t_{0}}^{+\infty}dt_2\int_{t_0}^{+\infty}dt_1\phi({t_2})\xi(t_1)\hat{b}^{\dagger}({t_2})\hat{b}^{\dagger}({t_{1}})\ket{vac},
\end{equation}
where
\begin{equation}
 \hat{b}({t})=\frac{1}{\sqrt{2\pi}}\int_{-\infty}^{+\infty}{ d}\omega\;\hat{b}({\omega}){e}^{-i\omega t},
 \end{equation}
and  $\mathcal{N}_{\xi\phi}$ is the normalization factor given by (\ref{factor}). 
The field operators in the time domain satisfy the following commutation relations 
\begin{equation}
[\hat{b}({t}),\hat{b}({t^{\prime}})]=0,\;\;[\hat{b}^{\dagger}({t}),\hat{b}^{\dagger}({t^{\prime}})]=0,\;\;[\hat{b}({t}),\hat{b}^{\dagger}({t^{\prime}})]=\delta(t-t^{\prime}).
\end{equation}
Note that only the symmetric part of the subintegral function gives a non-zero contribution to (\ref{tps1}) thus
the state can be rewritten as 
\begin{equation}
\ket{2_{\xi\phi}} =\frac{1}{2\sqrt{\mathcal{N}_{\xi\phi}}}\int_{t_0}^{+\infty}dt_2\int_{t_0}^{+\infty}dt_1\left(\phi({t_2})\xi(t_1)+\xi({t_2})\phi(t_2)\right)\hat{b}^{\dagger}({t_2})\hat{b}^{\dagger}({t_{1}})\ket{vac}.
\end{equation}
The Hamiltonian of the three-level system has the form
 \begin{equation}\label{Hamiltonian}
 \hat{H}_{\mathcal{S}}=-\omega_{eg}\ketbra{g}{g}+ \omega_{fe}\ketbra{f}{f}
 \end{equation}
 where $\omega_{eg}$ and  $\omega_{fe}$ are the transition frequencies of the atom.  
Suppose that the interaction of the atom with the unidirectional electromagnetic fields is characterized  by the coupling operator
\begin{equation}\label{coupling}
 \hat{L}=\sqrt{\Gamma_{e}}\ketbra{g}{e}+\sqrt{\Gamma_{f}}\ketbra{e}{f},
 \end{equation} 
 where $\Gamma_e$, $\Gamma_{f}$ are the positive coupling constants. The quantities 
$\Gamma_{f}^{-1}$ and $\Gamma_{e}^{-1}$ represent the lifetimes of states $\ket{f}$ and $\ket{e}$, respectively, due to spontaneous emission to the considered light mode. Taking the coupling operator in the form of (\ref{coupling}) means that the transition frequencies are close to each other, allowing both atomic transitions to be excited by the same field. 
 
 Making use of the quantum trajectories one can easily determine the formula for the probability that the atom is in the excited state $\ket{f}$ at time $t$. This probability can be written in the form $P_{f}(t)=\left\vert\left\vert\ket{\psi_{t|0}(0)}\right\vert\right\vert^2$, where $\ket{\psi_{t|0}(0)}$ is the conditional vector associated with the scenario that we do not observe any photon up to $t$ and there are no photons in the input field after $t$, which means that both photons of the input field have been absorbed by the atom. Thus, $P_{f}(t)$ is the two-photon absorption probability and for the state (\ref{tps1}) we obtain the formula 
\begin{equation}
P_{f}(t)=\frac{1}{\mathcal{N}_{\xi\phi}}\left\vert\bra{f}\hat{T}_{t}\int_{t_0}^{t}ds^{\prime}\int_{t_0}^{s^{\prime}}ds\left(\hat{A}_{s^{\prime}}(\xi)\hat{A}_{s}(\phi)+\hat{A}_{s^{\prime}}(\phi)\hat{A}_{s}(\xi)\right)\hat{T}_{-t_0}\ket{g}\right\vert^2, 
\end{equation}
where
\begin{equation}
\hat{A}_{t}(\xi)= -\xi(t)\left[\sqrt{\Gamma_e}e^{\left(i\omega_{eg}+\frac{\Gamma_e}{2}\right)t}\ketbra{e}{g}+\sqrt{\Gamma_f}e^{\left(i\omega_{fe}+\frac{1}{2}\left(\Gamma_f-\Gamma_e\right)\right)t}\ketbra{f}{e}\right],
\end{equation}
\begin{equation}
\hat{A}_{t}(\phi)= -\phi(t)\left[\sqrt{\Gamma_e}e^{\left(i\omega_{eg}+\frac{\Gamma_e}{2}\right)t}\ketbra{e}{g}+\sqrt{\Gamma_f}e^{\left(i\omega_{fe}+\frac{1}{2}\left(\Gamma_f-\Gamma_e\right)\right)t}\ketbra{f}{e}\right].
\end{equation}
Thus the probability of two-photon absorption for the unidirectional field prepared in the two-photon state (\ref{tps1}) is given by 
\begin{equation}\label{tpa1}
P_{f}(t)= \frac{\Gamma_e\Gamma_f e^{-\Gamma_ f t}}{\mathcal{N}_{\xi\phi}} \left|\int_{t_0}^{t}ds^{\prime}e^{\left(i\omega_{fe}+\frac{1}{2}(\Gamma_f-\Gamma_e)\right)s^{\prime}}\int_{t_0}^{s^{\prime}}ds e^{\left(i\omega_{eg} +\frac{\Gamma_e}{2}\right)s}
\left(\phi({s^{\prime}})\xi({s})+\xi(s^{\prime})\phi({s})\right)\right|^2.
\end{equation}
The above result can be generalized to the case when the input field is taken in the two-photon state vector of the form
\begin{equation}\label{tps2}
\ket{2_{\Phi}} =\frac{1}{\sqrt{\mathcal{N}_{\Phi}}}\int_{t_0}^{+\infty}dt_2\int_{t_0}^{+\infty}dt_1\Phi(t_2,t_1)\hat{b}^{\dagger}(t_2)\hat{b}^{\dagger}(t_{1})\ket{vac},
\end{equation}
where 
\begin{equation}
\mathcal{N}_{\Phi}=\int_{t_0}^{\infty}dt_2\int_{t_0}^{+\infty}dt_1\left(\vert\Phi(t_2,t_1)\vert^{2}+\Phi^{\ast}(t_1,t_2)\Phi(t_2,t_1)\right).
\end{equation}
Let us notice that such state can be rewritten as 
\begin{equation}\label{tps3}
\ket{2_{\Phi}} =\int_{t_0}^{+\infty}dt_2\int_{t_0}^{+\infty}dt_1\Phi_{sym}(t_2,t_1)\hat{b}^{\dagger}(t_2)\hat{b}^{\dagger}(t_{1})\ket{vac},
\end{equation}
where 
\begin{equation}
\Phi_{sym}(t_2,t_1)=\frac{1}{2\sqrt{\mathcal{N}_{\Phi}}}
\left(\Phi(t_2,t_1)+\Phi(t_1,t_2)\right).
\end{equation}
Using the linear decomposition of the two-photon state (\ref{tps3}) into separable two-photon states and referring to the linearity of the evolution equation for the total system, we obtain the formula for the two-photon absorption as follows
\begin{align}\label{tpa2}
P_{f}(t)&= \frac{\Gamma_e\Gamma_f e^{-\Gamma_{f}t}}{\mathcal{N}_{\Phi}} \left|\int_{t
_0}^{t}ds^{\prime}e^{\left(i\omega_{fe}+\frac{1}{2}(\Gamma_{f}-\Gamma_{e})\right)s^{\prime}}\int_{t_0}^{s^{\prime}}ds e^{(i\omega_{eg} +\frac{\Gamma_e}{2})s}
\left(\Phi(s^{\prime},s)+\Phi(s,s^{\prime})\right)\right|^2\nonumber\\
& = 4\Gamma_e\Gamma_fe^{-\Gamma_{f}t}\left|\int_{t_0}^{t}ds^{\prime}e^{\left(i\omega_{fe}+\frac{1}{2}(\Gamma_{f}-\Gamma_{e})\right)s^{\prime}}\int_{t_0}^{s^{\prime}}ds\; e^{\left(i\omega_{eg} +\frac{\Gamma_e}{2}\right)s}
\Phi_{sym}(s^{\prime},s)\right|^2.
\end{align}
The generalization of this result to the case where the system also interacts with an additional environment in the vacuum state is straightforward; it suffices to appropriately modify the coefficients in the exponents.

The derived formula for two-photon absorption can serve as a starting point for detailed studies on the optimal excitation of an atom by two-photon light. We will analyze in this paper the problem of optimal choice of the two-photon state for obtaining the maximum value of the excitation probability $P_{f}(t)$ when the interaction starts at time $t_0$. 
\begin{Theorem}\label{TH-3}
Maximal value of the probability $P_{f}(t)$ at time $t>t_{0}$ reads as follows
\begin{equation}\label{P-max1}
P^{\rm max}_{f}(t) = 1- \frac{1}{\Gamma_f-\Gamma_{e}}\left(\Gamma_{f}e^{-\Gamma_{e}(t-t_{0})} - \Gamma_{e}e^{-\Gamma_{f}(t-t_{0}) }\right),
\end{equation}
and is realized for the two-photon state of the profile 
\begin{align}\label{xi-max1}
 \Phi_{opt}(t_2,t_{1})=\frac{1}{\sqrt{\mathcal{N}_{opt}}} &\left(e^{\frac{1}{2}\Gamma_f t_2-\frac{1}{2}\Gamma_e(t_2-t_{1})} e^{-i\omega_{fe}t_2-i\omega_{eg}t_1} \chi_{t_0<t_1<t_2<t}\right.\nonumber\\&\left.+e^{\frac{1}{2}\Gamma_f t_1-\frac{1}{2}\Gamma_e(t_1-t_{2})} e^{-i\omega_{fe}t_1-i\omega_{eg}t_2}  \chi_{t_0<t_2<t_1<t}\right),
\end{align}
where $\chi$ is the 
characteristic function of a given subset of 
$\mathbb{R}^2$ for the variables $(t_2,t_1)$
and the normalization factor has the form
\begin{equation}\label{normopt1}
\mathcal{N}_{opt}=\frac{4e^{\Gamma_f t}}{\Gamma_{e}\Gamma_{f}}\left[1-e^{-\Gamma_f (t-t_0)}+\frac{\Gamma_{f}}{\Gamma_{e}-\Gamma_{f}}\left(e^{-\Gamma_{e}(t-t_0)}-e^{-\Gamma_{f}(t-t_0)}\right)\right].
\end{equation}
Let us clarify that $\chi_{t_0<t_1<t_2<t}$ is the characteristic function defined for the set of points satisfying $t_0<t_1<t_2<t$, i.e.,  a triangle spanned by vertices $(t_0,t_0)$, $(t,t_0)$, $(t,t)$ on the plane of variables $(t_2,t_1)$.
\end{Theorem}
The proof of Theorem \ref{TH-3} is provided in \ref{ProfTH-3}.  
\begin{remark}
        If $\Gamma_f = \Gamma_e=\Gamma$, the formula (\ref{P-max1}) reduces to:
        \begin{equation}
            1 - e^{-\Gamma (t-t_0)} (1 + \Gamma (t-t_0)).
        \end{equation}
    \end{remark}

\begin{figure}\label{fig:Phi_sym}
\begin{center}
\includegraphics[width=.75\textwidth]
{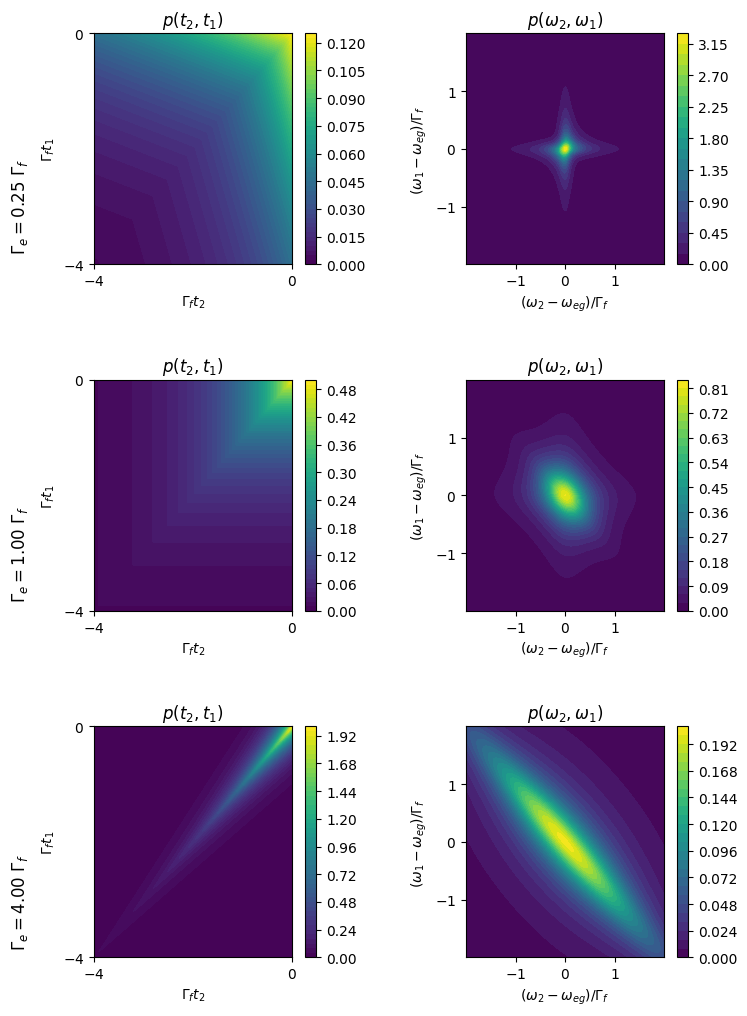}
\caption{Probability density functions in the frequency and time domains of the optimal two-photon state for the unidirectional field and the ratios $\Gamma_e/\Gamma_{f}$: $0.25$, $1$, and $4$ and $t=0$. See \url{https://github.com/gniewko-s/Two_Photon_Optimal_Input/blob/main/uni.py} for the interactive graph.}
\end{center}
\end{figure}

    \begin{remark}
If $t_{0}\to -\infty$, then we obtain the optimal two-photon state for the unidirectional field of the form
\begin{equation}\label{optstate1}
\sqrt{\Gamma_e\Gamma_f}e^{-\frac{\Gamma_f t}{2}}\int_{-\infty}^{t}dt_2\int_{-\infty}^{t_2}dt_1e^{\frac{1}{2}(\Gamma_f-\Gamma_{e})t_2}e^{\frac{\Gamma_e}{2}t_1}e^{-i\omega_{fe}t_2-i\omega_{eg}t_1} \hat{b}^{\dagger}(t_2)\hat{b}^{\dagger}(t_1)\ket{vac},
\end{equation}
which gives the probability $P_{f}(t)=1$. The state (\ref{optstate1}) in  the frequency domain is 
\begin{equation}
\int_{-\infty}^{+\infty}d\omega_2\int_{-\infty}^{+\infty}d\omega_1\tilde{\Phi}_{opt}(\omega_2,\omega_1)\hat{b}^{\dagger}(\omega_2)\hat{b}^{\dagger}(\omega_1)\ket{vac},
\end{equation}
where 
\begin{equation}
\tilde{\Phi}_{opt}(\omega_2,\omega_1)=\frac{\sqrt{\Gamma_e\Gamma_{f}}e^{i(\omega_1+\omega_2-\omega_{fg})t}}{4\pi\big[i\left(\omega_1+\omega_{2}-\omega_{fg}\right)+\frac{\Gamma_f}{2}\big]}\left[ \frac{1}{i\left(\omega_1-\omega_{eg}\right)+\frac{\Gamma_e}{2}}+\frac{1}{i\left(\omega_2-\omega_{eg}\right)+\frac{\Gamma_e}{2}}\right],
\end{equation}
where $\omega_{fg}=\omega_{fe}+\omega_{eg}$.
    \end{remark}
\begin{remark}
The two-photon state (\ref{optstate1}) is a time-reversed of the two-photon state emitted spontaneously by the three-level atom in a ladder configuration prepared initially in the state $\ket{f}$ for the case of unidirectional field.
\end{remark}
Although the experimental realization  of the optimal state is challenging, it can serve as a reference point in discussions on optimizing two-photon excitation using other two-photon states. Note that, in fact, there exist infinitely many optimal states, i.e., for any given time $t$, one can determine a two-photon state that leads to perfect excitation of the system at time $t$. 
The probability distributions of the optimal state (\ref{optstate1}) in the time and frequency domains for the case when $t=0$ are shown in Figure \ref{fig:Phi_sym}. We use a scale determined by $\Gamma_{f}^{-1}$ that is the lifetime of state $\ket{f}$ due to spontaneous emission. As shown in the illustration, the properties of the optimal state depend on the relationship between the lifetimes of states $\ket{e}$
 and $\ket{f}$. The figure presents the probability distributions for three different values of the ratio $\Gamma_e/\Gamma_f$: $0.25$, $1$, and $4$. The figure illustrates an example when $\omega_{fg}=2\omega_{eg}$.
Note that then the local maxima of the function in the frequency domain align along lines given by the equations:  $\omega_1=\omega_{eg}$, $\omega_{2}=\omega_{eg}$, and $\omega_1+\omega_2=2 \omega_{eg}$.  
 Thus one can observe the role of single-photon and double-photon resonances in the optimal excitation of the atom. As can be seen, their significance depends on the ratio $\Gamma_{e}/\Gamma_{f}$. Note that the symmetry of the probability distributions is related to the fact that the photons are indistinguishable in this case. The marginal probability distributions in time and frequency are identical. It is important to emphasize also that the photons in the optimal state are entangled.
Thus the entanglement plays an important role in the excitation of the three-level system. 

\subsection{Results for a bidirectional field in two-photons state}

In this section, we describe the results for the bidirectional field prepared in the two-photon state \cite{Loudon2000, RMS07}
\begin{equation}
\ket{1_{\xi},1_{\phi}} =\int_{t_0}^{+\infty}dt_2\int_{t_0}^{+\infty}dt_1\phi({t_2})\xi(t_1)\hat{b}_{2}^{\dagger}({t_2})\hat{b}^{\dagger}_{1}({t_{1}})\ket{vac,vac},
\end{equation}
where
\begin{equation}
 \hat{b}_{l}(t)=\frac{1}{\sqrt{2\pi}}\int_{-\infty}^{+\infty}d\omega\hat{b}_{l}(\omega){e}^{-i\omega t},\;\;\; l=1,2. 
 \end{equation}
The field operators in the time domain satisfy the commutation relations
\begin{equation}
[\hat{b}_{l}({t}),\hat{b}_{l^{\prime}}({t^{\prime}})]=0,\;\;[\hat{b}_{l}^{\dagger}({t}),\hat{b}_{l^{\prime}}^{\dagger}({t^{\prime}})]=0,\;\;[\hat{b}_{l}({t}),\hat{b}^{\dagger}_{l^{\prime}}({t^{\prime}})]=\delta_{ll^{\prime}}\delta(t-t^{\prime}).
\end{equation}
We assume that we are dealing now with the field of the two distinct central frequencies fitting to two different atomic transitions. The Hamiltonian of the atom has the form of (\ref{Hamiltonian}) and the interaction of the atom with the two unidirectional electromagnetic fields is characterized by the coupling operators:
 \begin{equation}
 \hat{L}_1=\sqrt{\Gamma_{e}}\ketbra{g}{e},\;\;\hat{L}_2=\sqrt{\Gamma_{f}}\ketbra{e}{f},
 \end{equation} 
 where $\Gamma_e$ and $\Gamma_f$ are non-negative coupling constants.

Using the fact that for the atom prepared in the ground state the probability of the excitation of the state $\ket{f}$ at time $t$ can be written in the form $P_{f}(t)=\left\vert\left\vert\ket{\psi_{t|0}(00)}\right\vert\right\vert^2$, where $\ket{\psi_{t|0}(00)}$ is  the conditional vector given by the formula (\ref{convec}), we obtain the formula
\begin{equation}\label{ptpa2}
P_{f}(t)=\left\vert\bra{f}\hat{T}_{t}\int_{t_0}^{t}ds^{\prime}\int_{t_0}^{s^{\prime}}ds\left(\hat{A}_{s^{\prime}}(\xi)\hat{A}_{s}(\phi)+\hat{A}_{s^{\prime}}(\phi)\hat{A}_{s}(\xi)\right)\ket{g}\right\vert^2,
\end{equation}
where the operators associated with the absorption of photons have the form
\begin{equation}
\hat{A}_{t}(\xi)= -\sqrt{\Gamma_e}\xi(t)e^{\left(i\omega_{eg}+\frac{\Gamma_e}{2}\right)t}\ketbra{e}{g},
\end{equation}
\begin{equation}
\hat{A}_{t}(\phi)= -\sqrt{\Gamma_f}\phi(t)e^{\left(i\omega_{fe}+\frac{1}{2}\left(\Gamma_f-\Gamma_e\right)\right)t}\ketbra{f}{e}.
\end{equation}
Hence the probability of the two-photon absorption is 
\begin{equation}\label{tpa3}
P_{f}(t)= \Gamma_e\Gamma_f \left|\int_{t_0}^{t}ds^{\prime}\phi(s^{\prime})e^{-\left(i\omega_{fe}+\frac{\Gamma_f}{2}\right)\left(t-s^{\prime}\right)}\int_{t_0}^{s^{\prime}}ds\xi(s)e^{i\omega_{eg} s-\frac{\Gamma_e}{2}(s^{\prime}-s)}\right|^2.
\end{equation}
In a more general case, the two-photon state field is defined by 
\begin{equation}\label{tps4}
\ket{2_{\Phi}} =\int_{t_0}^{+\infty}dt_2\int_{t_0}^{+\infty}dt_1\Phi(t_2,t_1)\hat{b}_{2}^{\dagger}(t_2)\hat{b}_{1}^{\dagger}(t_{1})\ket{vac,vac},
\end{equation}
where
\begin{equation}
\int_{t_0}^{\infty}dt_2\int_{t_0}^{+\infty}dt_1\vert\Phi(t_2,t_1)\vert^2=1. 
\end{equation}
By referring to the similar arguments to those presented in the previous subsection, which allowed us to determine formula (\ref{tpa2}), we obtain  for the state (\ref{tps4}), the probability of the two-photon absorption of the form
\begin{equation}\label{tpa4}
P_{f}(t)= \Gamma_e\Gamma_f \left|\int_{t_0}^{t}ds^{\prime}e^{-\left(i\omega_{fe}+\frac{\Gamma_f}{2}\right)\left(t-s^{\prime}\right)}\int_{t_0}^{s^{\prime}}dse^{i\omega_{eg} s-\frac{\Gamma_e}{2}(s^{\prime}-s)}\Phi(s^{\prime},s)\right|^2.
\end{equation}

Below we shall present an analysis of the optimal excitation of the atom by the bidirectional field.  

\begin{Theorem}\label{TH-4}
Maximal value of the probability $P_{f}(t)$ at time $t>t_{0}$ reads as follows
\begin{equation}\label{P-max}
P^{\rm max}_{f}(t) = 1- \frac{1}{\Gamma_f-\Gamma_{e}}\left(\Gamma_{f}e^{-\Gamma_{e}(t-t_{0})} - \Gamma_{e}e^{-\Gamma_{f}(t-t_{0}) }\right),
\end{equation}
and is realized for the two-photon state of the profile 
\begin{equation}\label{xi-max}
 \Phi_{opt}(t_2,t_{1})=\frac{1}{\sqrt{\mathcal{N}_{opt}}} e^{\frac{1}{2}(\Gamma_f-\Gamma_e)t_2}e^{\frac{\Gamma_{e}t_{1}}{2}}e^{-i\omega_{fe}t_2-i\omega_{eg}t_1}  \chi_{t_0<t_1<t_2<t}. 
\end{equation}
 Here the normalization factor has the form
\begin{equation}\label{normopt}
\mathcal{N}_{opt}=\frac{e^{\Gamma_f t}}{\Gamma_{e}\Gamma_{f}}\left[1-e^{-\Gamma_f (t-t_0)}+\frac{\Gamma_{f}}{\Gamma_{e}-\Gamma_{f}}\left(e^{-\Gamma_{e}(t-t_0)}-e^{-\Gamma_{f}(t-t_0)}\right)\right].
\end{equation}
\end{Theorem}
The proof of Theorem \ref{TH-4} is given  \ref{ProfTH-4}.  
    \begin{remark}
If $t_{0}\to -\infty$, then we obtain for the bidirectional field the optimal two-photon state of the form
\begin{equation}\label{optstate}
\sqrt{\Gamma_e\Gamma_f}e^{-\frac{\Gamma_f t}{2}}\int_{-\infty}^{t}dt_2\int_{-\infty}^{t_2}dt_1e^{\frac{1}{2}(\Gamma_f-\Gamma_{e})t_2}e^{\frac{\Gamma_e}{2}t_1}e^{-i\omega_{fe}t_2-i\omega_{eg}t_1} \hat{b}^{\dagger}_{2}(t_2)\hat{b}^{\dagger}_{1}(t_1)\ket{vac,vac},
\end{equation}
which gives the probability $P_{f}(t)=1$. The state (\ref{optstate}) in  the frequency domain is 
\begin{equation}
\int_{-\infty}^{+\infty}d\omega_2\int_{-\infty}^{+\infty}d\omega_1 \tilde{\Phi}_{opt}(\omega_2,\omega_1)\hat{b}^{\dagger}_2(\omega_2)\hat{b}^{\dagger}_1(\omega_1)\ket{vac,vac},
\end{equation}
where 
\begin{equation}
\tilde{\Phi}_{opt}(\omega_2,\omega_1)=\frac{\sqrt{\Gamma_e\Gamma_{f}}e^{i(\omega_1+\omega_2-\omega_{fg})t}}{2\pi\big[ i\left(\omega_1-\omega_{eg}\right)+\frac{\Gamma_e}{2}\big]\big[i\left(\omega_1+\omega_{2}-\omega_{fg}\right)+\frac{\Gamma_f}{2}\big]}.
\end{equation}

    \end{remark}

    \begin{remark}
The two-photon state (\ref{optstate}) is the time-reversed to the two-photon state of light emitted spontaneously by the three-level atom in a ladder configuration prepared initially in the state $\ket{f}$ for the case of the bidirectional field.
  \end{remark}

The description of two-photon spontaneous emission for a three-level atom in a ladder configuration, interacting with a vacuum field and initially prepared in the state $\ket{f}$, can be found in \cite{Scully1997}. The authors considered a multi-directional vacuum input field and determined the multi-directional two-photon state of the field emitted by the atom. 

\begin{figure}\label{fig:Phi}
\begin{center}
\includegraphics[width=.75\textwidth]
{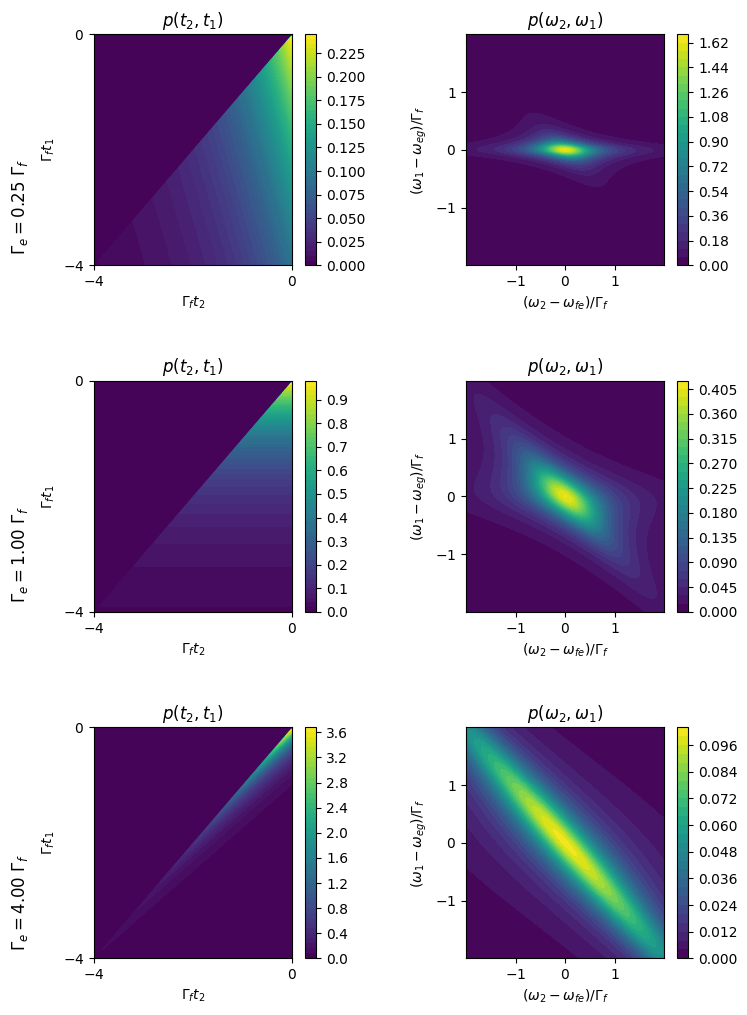}
\caption{Probability density functions in the frequency and time domains of the optimal two-photon state for the bidirectional field and the ratios $\Gamma_e/\Gamma_{f}$: $0.25$, $1$, and $4$ and $t=0$.
See \url{https://github.com/gniewko-s/Two_Photon_Optimal_Input/blob/main/bi.py} for interactive graph.}
\end{center}
\end{figure}

The probability distributions of the optimal state (\ref{optstate}) in the time and frequency domains are shown in Figure \ref{fig:Phi}. Here, similarly to the previous case, we use the reference scale given by the lifetime $\Gamma_{f}^{-1}$ of the excited state $\ket{f}$  due to spontaneous emission and we illustrate the situation where $P_{f}(t=0)=1$. The figure presents the probability distributions for three different values of the ratio $\Gamma_e/\Gamma_f$: $0.25$, $1$, and $4$. For the probability distributions in the frequency domain, the local maxima align along the lines given by the equations: $\omega_1=\omega_{eg}$, $\omega_{2}=\omega_{fe}$, and $\omega_1+\omega_2=\omega_{fg}$.  They correspond to single resonances and double resonance. It is worth emphasizing that, in this case, unlike for the unidirectional field, the probability densities are not symmetric functions. Physically, this is related to the fact that the photons are distinguishable. Their marginal distributions in time and frequency are different. In this case, similarly to the unidirectional field, the photons in the optimal state are entangled.
 
 \section{Conclusions}

In this paper, we have investigated the problem of quantum trajectories for a quantum system interacting with a propagating field prepared in a two-photon state. We have considered two types of the system environment: a unidirectional field and a bidirectional field. Due to temporal correlations of the input field, the evolution of an open system is non-Markovian in this case and given by the sets of coupled filtering or master equations \cite{Baragiola12,Dong15,Zhang2016,Dabrowska2019b,Zhang2019b}.
The evolution of the system interacting with a traveling field with a definite number of photons is quite complicated, particularly if the photons have different profiles or, even more so, if the photons are entangled. We have determined general analytical formulae for quantum trajectories associated with stochastic counting processes. We have formulated and solved this problem using a collision model. This approach has allowed us to provide a rigorous description of indirect quantum measurement performed on the output field and to the conditional evolution of the quantum system. 
The recurrence equations for the conditional vectors associated with the counting measurement give us an insight into the problem of excitation of the system and reveal the properties of the output field. We have obtained the recipe for the {\it a posteriori} state of an arbitrary quantum system.  Using these formulae, one can determine both the conditional as well as unconditional evolution of the open system, and describe the properties of the output field. Our formulae for quantum trajectories can be used to a variety of open systems, including multi-level atoms and resonant cavities.

We have applied the quantum trajectories to derive the formula for the probability of two-photon absorption for a three-level atom in a ladder configuration. The problem of two-photon absorption for propagating light in a two-photon state with a pair of entangled photons is currently being intensively studied by both theorists \cite{Schlawin2017,Schlawin2021,Raymer2021} and experimental physicists \cite{Thew2021,Thew2022}. We have obtained the analytical formulae describing the excitation of the atom by unidirectional and bidirectional light for an arbitrary two-photon state vector. Thus, we have also shown that even in the general case, when the number of equations defining the problem becomes infinite, it is still possible to obtain analytical results. We have also determined the optimal two-photon state that gives a perfect excitation of the atom. It is worth noting that the optimal state is time-reversed to a two-photon state emitted spontaneously by a three-level atom with a ladder configuration prepared initially at the highest energy level state. The generalisation of these results to the case of a multi-directional field is rather straightforward.  An analogous result for the perfect excitation of a two-level atom by the single-photon wavepacket with the rising exponential profile was given in  \cite{WMSS11,Stobinska10a}, and examined in the laboratory \cite{Scarani2013,Leong2016}. We have verified that single-photon and double-photon resonances play a crucial role in the optimal excitation of a three-level system. The photons in the optimally exciting state of the three-level system are entangled. The derived formulas for the two-photon excitation probability can serve as a starting point for further research on the role of entanglement and interference in the excitation process of a three-level system by two-photon light.

\section*{Acknowledgements}

\noindent {\bf Funding information} This research was supported partially by the National Science Centre project (Narodowe Centrum Nauki) 2018/30/A/ST2/00837. This project was also funded within the QuantERA II Programme that has received funding from the European Union’s Horizon 2020 research and innovation programme under Grant Agreement No 101017733, and with The National Centre for Research and Development, Poland, grant number QUANTERAII/1/21/E2TPA/2023.

\appendix

\section{Proof of Theorem \ref{TH-1} } \label{ProofTH-1}

We prove Theorem \ref{TH-1} using the induction method. Assuming that (\ref{condTH1}) is true for $j\in \mathbb{N}$, we can use the property (\ref{prop1}) to observe that 
\begin{align}
\ket{\Psi_{j}}=&
\ket{0}_{j}\otimes\bigg[\ket{vac}_{[j+1}\otimes \ket{\psi_{j}(0)}+\ket{1_{\xi}}_{[j+1}\otimes \ket{\psi_{j}(1_{\xi})}\nonumber\\
&+\ket{1_{\phi}}_{[j+1}\otimes \ket{\psi_{j}(1_{\phi})}+\ket{2_{\xi\phi}}_{[j+1}\otimes\ket{\psi_{j}(2_{\xi\phi})}\bigg]\nonumber\\
&+\ket{1}_{j}\otimes 
\bigg[\ket{vac}_{[j+1}\otimes \left( \sqrt{\tau}\xi_{j}\ket{\psi_{j}(1_{\xi})}+\sqrt{\tau}\phi_{j}\ket{\psi_{j}(1_{\phi})}\right)\nonumber\\
&+\ket{1_{\xi}}_{[j+1}\otimes \frac{\sqrt{\tau}\phi_{j}}{\sqrt{\mathcal{N}_{\xi\phi}}}\ket{\psi_{j}(2_{\xi\phi})}+
\ket{1_{\phi}}_{[j+1}\otimes \frac{\sqrt{\tau}\xi_{j}}{\sqrt{\mathcal{N}_{\xi\phi}}}\ket{\psi_{j}(2_{\xi\phi})}\bigg]\nonumber\\
&+\ket{2}_{j}\otimes\ket{vac}_{[j+1}\otimes \frac{\sqrt{2}\tau\xi_{j}\phi_{j}}{\sqrt{\mathcal{N}_{\xi\phi}}}\ket{\psi_{j}(2_{\xi\phi})}.
\end{align}
The unitary operator $\hat{\mathbb{V}}_{[j}$ acting on the vector $\ket{\Psi_{j}}$ gives
\begin{align}
\hat{\mathbb{V}}_{[j} \,\ket{\Psi_{j}}=& \sum_{m=0}^{+\infty}\ket{n}_{j}\otimes\bigg[
\ket{vac}_{[j+1}\otimes \bigg( \hat{V}_{m0}\ket{\psi_{j}(0)}+\sqrt{\tau}\xi_{j}\hat{V}_{m1}\ket{\psi_{j}(1_{\xi})}\nonumber\\
&+\sqrt{\tau}\phi_{j}\hat{V}_{m1}\ket{\psi_{j}(1_{\phi})}+\frac{\sqrt{2}\tau\xi_{j}\phi_{j}}{\sqrt{\mathcal{N}_{\xi\phi}}}\hat{V}_{m2}\ket{\psi_{j}(2_{\xi\phi})}\bigg)\nonumber\\
&+\ket{1_{\xi}}_{[j+1}\otimes\bigg(\hat{V}_{m0}\ket{\psi_{j}(1_{\xi})}+\frac{\sqrt{\tau}\phi_{j}}{\sqrt{\mathcal{N}_{\xi\phi}}}\hat{V}_{m1}\ket{\psi_{j}(2_{\xi\phi})}\bigg)\nonumber\\
&+\ket{1_{\phi}}_{[j+1}\otimes\bigg(\hat{V}_{m0}\ket{\psi_{j}(1_{\phi})}+\frac{\sqrt{\tau}\xi_{j}}{\sqrt{\mathcal{N}_{\xi\phi}}}\hat{V}_{m1}\ket{\psi_{j}(2_{\xi\phi})}\bigg)\nonumber\\
&+\ket{2_{\xi\phi}}_{[j+1}\otimes \hat{V}_{m0}\ket{\psi_{j}(2_{\xi\phi})}\bigg]
\end{align}
The conditional vector $\ket{\Psi_{j+1}}$ from  $\mathcal{H}_{\mathcal{E}}^{[j+1}\otimes \mathcal{H}_{S}$ is defined by
\begin{equation}
\Pi_{\eta_{j+1}}\otimes\hat{\mathbbm{1}}_{\mathcal{E}}^{[j+1}\otimes\hat{\mathbbm{1}}_{\mathcal{S}}\,\hat{\mathbb{V}}_{[j}\,\ket{\Psi_{j}}=\ket{\eta_{j+1}}_{j}\otimes\ket{\Psi_{j+1}},\nonumber
\end{equation}
where $\Pi_{\eta_{j+1}}=|\eta_{j+1}\rangle_{j}{}_{j}\langle \eta_{j+1}|$, and $\eta_{j+1}\in \mathbb{N}$ is the result of the measurement of (\ref{obs}) at the time $\tau(j+1)$. Hence
\begin{align}\label{cond}
\ket{\Psi_{j+1}}=& 
\ket{vac}_{[j+1}\otimes \bigg( \hat{V}_{\eta_{j+1}0}\ket{\psi_{j}(0)}+\sqrt{\tau}\xi_{j}\hat{V}_{\eta_{j+1}1}\ket{\psi_{j}(1_{\xi})}
\nonumber\\
&+\sqrt{\tau}\phi_{j}\hat{V}_{\eta_{j+1}1}\ket{\psi_{j}(1_{\phi})}+\frac{\sqrt{2}\tau\xi_{j}\phi_{j}}{\sqrt{\mathcal{N}_{\xi\phi}}}\hat{V}_{\eta_{j+1}2}\ket{\psi_{j}(2_{\xi\phi})}\bigg)\nonumber\\
&+\ket{1_{\xi}}_{[j+1}\otimes\bigg(\hat{V}_{\eta_{j+1}0}\ket{\psi_{j}(1_{\xi})}+\frac{\sqrt{\tau}\phi_{j}}{\sqrt{\mathcal{N}_{\xi\phi}}}\hat{V}_{\eta_{j+1}1}\ket{\psi_{j}(2_{\xi\phi})}\bigg)\nonumber\\
&+\ket{1_{\phi}}_{[j+1}\otimes\bigg(\hat{V}_{\eta_{j+1}0}\ket{\psi_{j}(1_{\phi})}+\frac{\sqrt{\tau}\xi_{j}}{\sqrt{\mathcal{N}_{\xi\phi}}}\hat{V}_{\eta_{j+1}1}\ket{\psi_{j}(2_{\xi\phi})}\bigg)\nonumber\\
&+\ket{2_{\xi\phi}}_{[j+1}\otimes \hat{V}_{\eta_{j+1}0}\ket{\psi_{j}(2_{\xi\phi})},
\end{align}
and it end the proof.

\section{Generalization of Theorem \ref{TH-1}}  \label{B}

Let us consider the two-photon state vector defined in the Hilbert space $\mathcal{H}_{\mathcal{E}}=\bigotimes_{k=0}^{M-1}\mathcal{H}_{\mathcal{E},k}$ by the formula 
\begin{equation}\label{arbitrary}
\ket{2_{{\Phi}}}=\frac{1}{\sqrt{\mathcal{N}_{\Phi}}}\sum_{k_{1},k_{2}=0}^{M-1}
\tau \Phi_{k_{2}, k_{1}}\hat{\tilde{b}}_{k_{2}}^{\dagger}\hat{\tilde{b}}_{k_{1}}^{\dagger}\ket{vac},
\end{equation}
where $\mathcal{N}_{\Phi}$, given by (\ref{norm}), stands for the normalization factor.
Any state of this form can be decomposed as 
\begin{equation}
\ket{2_{\Phi}}=\sum_{n=0}^{M-1}u(n) \left(\hat{B}^{\dagger}[\xi(n)]\right)^2\ket{vac},
\end{equation}
where $\{\xi(n)\}_{n=0}^{M-1}$ is the set of orthogonal functions in $\mathcal{H}_{\mathcal{E}}$, i.e.,  $\sum_{k=0}^{M-1}\tau (\xi_{k}(n^{\prime}))^{\ast}\xi_{k}(n)=\delta_{nn^{\prime}}$), and $\sum_{n=0}^{M-1}u^{2}(n)=\frac{1}{2}$. We assume here that the phases have been incorporated into the definitions of the basis states and the coefficients are positive. 

\begin{Theorem}\label{TH-5} The conditional state vector of the input part of the environment (the part of the environment which has not interacted with $\mathcal{S}$ up to $j\tau$) and the system $\mathcal{S}$, for the initial state 
\begin{equation}
 \ket{2_{\Phi}}\otimes \ket{\psi_{0}}
\end{equation}
and the measurement of (\ref{obs}), at time $j\tau$ has the form
	\begin{equation}
	\ket{\tilde{\Psi}_{j}} 
	= \frac{\ket{\Psi_{j}}}
	{\sqrt{\bra{\Psi_{j}}\ket{\Psi_{j}}}},
	\end{equation}
	where $\ket{\Psi_{j}}$ is the unnormalized conditional vector from the Hilbert space $\mathcal{H}_\mathcal{E}^{[j} \otimes \mathcal{H}_\mathcal{S}$ such that
	\begin{align}\label{condTH3}
	\ket{\Psi_{j}}=&
	\ket{vac}_{[j}\otimes\sum_{n}u(n)\ket{\psi_{j}(0_{\xi(n)})}+
\sum_{n=0}^{+\infty}u(n)\ket{1_{\xi(n)}}_{[j}\otimes\ket{\psi_{j}(1_{\xi(n)})}\nonumber\\&+\sum_{n}u(n)\ket{2_{\xi(n)}}_{[j}\otimes\ket{\psi_{j}(2_{\xi(n)})},
	\end{align}
	where $\ket{2_{\xi(n)}}_{[j}= \left(\hat{B}^{\dagger}_{[j}[\xi(n)]\right)^2\ket{vac}_{[j}$ and the conditional vectors $\ket{\psi_{j}(0_{\xi({n})})}$, $\ket{\psi_{j}(1_{\xi(n)})}$, $\ket{\psi_{j}(2_{\xi({n})})}$ from the Hilbert space $\mathcal{H}_{\mathcal{S}}$ satisfy the set of coupled recurrence equations:
	\begin{align}\label{TH-3_rec1}
	\ket{\psi_{j+1}(0_{\xi({n})})} =& \hat{V}_{\eta_{j+1}0}\ket{\psi_{j}(0_{\xi({n})})}+\sqrt{\tau}\xi_{j}({n})\hat{V}_{\eta_{j+1}1}\ket{\psi_{j}(1_{\xi({n})})},\\
\ket{\psi_{j}(1_{\xi({n})})} =& \hat{V}_{\eta_{j+1}0}\ket{\psi_{j}(1_{\xi({n})})}+2\sqrt{\tau}\xi_{j}({n})\hat{V}_{\eta_{j+1}1}\ket{\psi_{j}(2_{\xi\phi})},\\
		\ket{\psi_{j+1}(2_{\xi({n})})} =& \hat{V}_{\eta_{j+1}0}\ket{\psi_{j}(2_{\xi({n})})},\label{TH-3_rec2}
	\end{align}
and initially we have $\ket{\psi_{j=0}(0_{\xi(n)})}=\ket{\psi_{j=0}(1_{\xi({n})})}=0$, $\ket{\psi_{j=0}(2_{\xi({n})})}=\ket{\psi_{0}}$. 

\end{Theorem}

\noindent The proof is analogous to the proof of Theorem 1. 

\section{The interaction operators}\label{Inter_Oper}

Writing down the unitary operator, $\exp\left(-i\tau \hat{H}_{k}\right)$ from Section \ref{BF} in the representation of the photon numbers  we get the following operators 
\begin{align}\label{vmatrix1}
\hat{V}_{00,00}&= \hat{\mathbbm{1}}-i\tau \left[\hat{H}_{\mathcal{S}}\!-\!  \frac{i}{2}\left(\hat{L}^\dagger_{1} \hat{L}_{1}+\hat{L}^\dagger_{2} \hat{L}_{2}\right) \right]+\hat{O}(\tau^{2}),\\
\hat{V}_{00,01}&=\hat{V}_{10,11}=-\sqrt{\tau} \hat{L}^{\dagger}_{2} + \hat{O}(\tau^{3/2}), \\
\hat{V}_{00,10}&=\hat{V}_{01,11}=- \sqrt{\tau} \hat{L}^\dagger_{1} + \hat{O}(\tau^{3/2}),\\
\hat{V}_{01,00}&= \hat{V}_{11,10}=\sqrt{\tau}\hat{L}_{2}+\hat{O}(\tau^{3/2}),\\
\hat{V}_{01,01}&= \hat{\mathbbm{1}}-i\tau \left[\hat{H}_{\mathcal{S}}- \frac{i}{2}\left(\hat{L}^\dagger_{1} \hat{L}_{1}+\hat{L}_{2} \hat{L}_{2}^{\dagger}\right) \right]+\hat{O}(\tau^{2}),\\
\hat{V}_{10,00}&=\hat{V}_{11,01}= \sqrt{\tau} \hat{L}_{1} + \hat{O}(\tau^{3/2}),\\
\hat{V}_{10,10}&=\hat{\mathbbm{1}}-i\tau \left[\hat{H}_{\mathcal{S}}-  \frac{i}{2}\left(\hat{L}_{1}\hat{L}_{1}^{\dagger}+\hat{L}_{2}^{\dagger}\hat{L}_{2}\right) \right]+\hat{O}(\tau^{2}),\\
\hat{V}_{11,11}&= \hat{\mathbbm{1}}-i\tau \left[\hat{H}_{\mathcal{S}}-  \frac{i}{2}\left(\hat{L}_{1} \hat{L}_{1}^{\dagger}+\hat{L}_{2}\hat{L}_{2}^{\dagger} \right) \right]+\hat{O}(\tau^{2}),\\
\hat{V}_{11,00}&=\frac{\tau}{2}\left(\hat{L}_{1}\hat{L}_{2}+\hat{L}_{2}\hat{L}_{1}\right)+\hat{O}(\tau^2),\\
\hat{V}_{00,11}&=\frac{\tau}{2}\left(\hat{L}_{1}^{\dagger}\hat{L}_{2}^{\dagger}+\hat{L}_{2}^{\dagger}\hat{L}_{1}^{\dagger}\right)+\hat{O}(\tau^2),\\
\hat{V}_{01,10}&=-\frac{\tau}{2}\left(\hat{L}_{1}^{\dagger}\hat{L}_{2}+\hat{L}_{2}\hat{L}_{1}^{\dagger}\right)+\hat{O}(\tau^2),\\
\hat{V}_{10,01}&=-\frac{\tau}{2}\left(\hat{L}_{1}\hat{L}_{2}^{\dagger}+\hat{L}_{2}^{\dagger}\hat{L}_{1}\right)+\hat{O}(\tau^2).\label{vmatrix2}
\end{align}

\section{Proof of Theorem \ref{TH-2} } \label{ProofTH-2} 

We prove Theorem \ref{TH-2} by the induction. In the first step, we check the initial conditions for $j=0$, then we assume that (\ref{TH-2cond4}) is true for $j\in \mathbb{N}$, and referring to (\ref{prop1}) we get  
\begin{align}
\ket{\Psi_{j}}=&
\ket{00}_{j}\otimes\bigg[\ket{vac}_{[j+1}\otimes\ket{vac}_{[j+1}\otimes \ket{\psi_{j}(00)}+\ket{1_{\xi}}_{[j+1}\otimes \ket{vac}_{[j+1}\otimes\ket{\psi_{j}(1_{\xi}0)}\nonumber\\
&+\ket{vac}_{[j+1}\otimes\ket{1_{\phi}}_{[j+1}\otimes \ket{\psi_{j}(01_{\phi})}+\ket{1_{\xi}}_{[j+1}\otimes\ket{1_{\phi}}_{[j+1}\otimes\ket{\psi_{j}(1_{\xi}1_{\phi})}\bigg]\nonumber\\
&+\ket{10}_{j}\otimes 
\bigg[\ket{vac}_{[j+1}\otimes \ket{vac}_{[j+1}\otimes\sqrt{\tau}\xi_{j}\ket{\psi_{j}(1_{\xi}0)}\nonumber\\
&+\ket{vac}_{[j+1}\otimes\ket{1_{\phi}}_{[j+1}\otimes\sqrt{\tau}\xi_{j}\ket{\psi_{j}(1_{\xi}1_{\phi})}\bigg]\nonumber\\
&+\ket{01}_{j}\otimes 
\bigg[\ket{vac}_{[j+1}\otimes \ket{vac}_{[j+1}\otimes\sqrt{\tau}\phi_{j}\ket{\psi_{j}(01_{\phi})}\nonumber\\
&+\ket{1_{\xi}}_{[j+1}\otimes\ket{vac}_{[j+1}\otimes\sqrt{\tau}\phi_{j}\ket{\psi_{j}(1_{\xi}1_{\phi})}\bigg]\nonumber\\
&+\ket{11}_{j}\otimes\ket{vac}_{[j+1}\otimes \ket{vac}_{[j+1}\otimes\tau\xi_{j}\phi_{j}\ket{\psi_{j}(1_{\xi}1_{\phi})}.
\end{align}
The unitary operator $\hat{\mathbb{V}}_{[j}$ acting on the vector $\ket{\Psi_{j}}$ gives
\begin{align}
\hat{\mathbb{V}}_{[j} \,\ket{\Psi_{j}}=& \sum_{n,n^{\prime}=0}^{+\infty}\ket{nn^{\prime}}_{j}\otimes\bigg\{
\ket{vac}_{[j+1}\otimes\ket{vac}_{[j+1}\otimes \bigg[\hat{V}_{nn^{\prime},00}\ket{\psi_{j}(00)}\nonumber\\
&+\sqrt{\tau}\xi_{j}\hat{V}_{nn^{\prime},10}\ket{\psi_{j}(1_{\xi}0)}
+\sqrt{\tau}\phi_{j}\hat{V}_{nn^{\prime},10}\ket{\psi_{j}(01_{\phi})}\nonumber\\&+\tau\xi_{j}\phi_{j}\hat{V}_{nn^{\prime},11}\ket{\psi_{j}(1_{\xi}1_{\phi})}\bigg]\nonumber\\
&+\ket{1_{\xi}}_{[j+1}\otimes\ket{vac}_{[j+1}\otimes\left(\hat{V}_{nn^{\prime},00}\ket{\psi_{j}(1_{\xi}0)}+{\sqrt{\tau}\phi_{j}}\hat{V}_{nn^{\prime},01}\ket{\psi_{j}(1_{\xi}1_{\phi})}\right)\nonumber\\
&+\ket{vac}_{[j+1}\otimes\ket{1_{\phi}}_{[j+1}\otimes\left(\hat{V}_{nn^{\prime},00}\ket{\psi_{j}(01_{\phi})}+{\sqrt{\tau}\xi_{j}}\hat{V}_{nn^{\prime},10}\ket{\psi_{j}(1_{\xi}1_{\phi})}\right)\nonumber\\
&+\ket{1_{\xi}}_{[j+1}\otimes \ket{1_{\xi}}_{[j+1}\otimes\hat{V}_{nn^{\prime},00}\ket{\psi_{j}(1_{\xi}1_{\phi})}\bigg\}
\end{align}
The conditional vector $\ket{\Psi_{j+1}}$ from $\mathcal{H}_{\mathcal{E}}^{[j+1}\otimes \mathcal{H}_{\mathcal{S}}$ is defined by
\begin{equation}
\Pi_{\eta_{j+1}}\otimes\hat{\mathbbm{1}}_{\mathcal{E}}^{[j+1}\otimes\hat{\mathbbm{1}}_{\mathcal{S}}\,\hat{\mathbb{V}}_{[j}\,\ket{\Psi_{j}}=\ket{\eta_{j+1}}_{j}\otimes\ket{\Psi_{j+1}},\nonumber
\end{equation}
where $\Pi_{\eta_{j+1}}=|\eta_{j+1}\rangle_{j}{}_{j}\langle \eta_{j+1}|$, and $\eta_{j+1}\in \mathbb{N}^2$ is the result of the measurement of (\ref{obs2}) at the time $\tau(j+1)$. Hence
\begin{align}
\ket{\Psi_{j+1}}=& \ket{vac}_{[j+1}\otimes\ket{vac}_{[j+1}\otimes\bigg[\hat{V}_{\eta_{j+1},00}\ket{\psi_{j}(00)}+\sqrt{\tau}\xi_{j}\hat{V}_{\eta_{j+1},10}\ket{\psi_{j}(1_{\xi}0)}\nonumber\\
&
+\sqrt{\tau}\phi_{j}\hat{V}_{\eta_{j+1},10}\ket{\psi_{j}(01_{\phi})}+\tau\xi_{j}\phi_{j}\hat{V}_{\eta_{j+1},11}\ket{\psi_{j}(1_{\xi}1_{\phi})}\bigg]\nonumber\\
&+\ket{1_{\xi}}_{[j+1}\otimes\ket{vac}_{[j+1}\otimes\left(\hat{V}_{\eta_{j+1},00}\ket{\psi_{j}(1_{\xi}0)}+{\sqrt{\tau}\phi_{j}}\hat{V}_{\eta_{j+1},01}\ket{\psi_{j}(1_{\xi}1_{\phi})}\right)\nonumber\\
&+\ket{vac}_{[j+1}\otimes\ket{1_{\phi}}_{[j+1}\otimes\left(\hat{V}_{\eta_{j+1},00}\ket{\psi_{j}(01_{\phi})}+{\sqrt{\tau}\xi_{j}}\hat{V}_{\eta_{j+1},10}\ket{\psi_{j}(1_{\xi}1_{\phi})}\right)\nonumber\\
&+\ket{1_{\xi}}_{[j+1}\otimes \ket{1_{\xi}}_{[j+1}\otimes\hat{V}_{\eta_{j+1},00}\ket{\psi_{j}(1_{\xi}1_{\phi})}
\end{align}
and it ends the proof.

\section{Generalization of Theorem \ref{TH-2}}  \label{GeneralizationTH-2}

Let us consider the two-photon state vector defined in the Hilbert space $\mathcal{H}_{\mathcal{E}}=\mathcal{H}_{\mathcal{E}_1}\otimes\mathcal{H}_{\mathcal{E}_2}$  by the formula (\ref{arbitrary2}). 
By applying the Schmidt decomposition any two-photon 
state of form (\ref{arbitrary2}) can be represented by the expression, 
\begin{equation}
\ket{2_{\Phi}}=\sum_{n=0}^{M-1}u(n) \ket{1_{\xi(n)}}\otimes \ket{1_{\phi(n)}},
\end{equation}
where $\sum_{k=0}^{M-1}\tau (\xi_{k}(n^{\prime}))^{\ast}\xi_{k}(n)=\delta_{nn^{\prime}}$, 
$\sum_{k=0}^{M-1}\tau (\phi_{k}(n^{\prime}))^{\ast}\phi_{k}(n)=\delta_{nn^{\prime}}$, and $\sum_{n=0}^{+\infty}u(n)^2=1$. 

\begin{Theorem}\label{TH-6} The conditional state vector of the input part of the environment (the part of the environment which has not interacted with $\mathcal{S}$ up to $j\tau$) and the system $\mathcal{S}$, for the initial state 
\begin{equation}
\sum_{n=0}^{+\infty}u(n) \ket{1_{\xi(n)}}\otimes \ket{1_{\phi(n)}}\otimes \ket{\psi_{0}}
\end{equation}
and the measurement of (\ref{obs2}), is at time $j\tau $ is given by 
	\begin{equation}\label{TH-6cond3}
	\ket{\tilde{\Psi}_{j}} = \frac{\ket{\Psi_{j}}}{\sqrt{\braket{\Psi_{j}}{\Psi_{j}}}},
	\end{equation}
	where $ \ket{\Psi_{j}}$ is the unnormalized conditional vector from the Hilbert space $\mathcal{H}_{\mathcal{E}}^{[j} \otimes \mathcal{H}_\mathcal{S}$ having the form
	\begin{align}\label{TH-6cond4}
	\ket{\Psi_{j}} =&\sum_{n=0}^{+\infty}u(n)\bigg[\ket{vac}_{[j}\otimes \ket{vac}_{[j}\otimes \ket{\psi_{j}(0_{\xi(n)}0_{\phi(n)})}+ \ket{1_{\xi(n)}}_{[j}\otimes\ket{vac}_{[j}\otimes \ket{\psi_{j}(1_{\xi(n)}0)} \nonumber\\ 
&+ \ket{vac}_{[j}\otimes \ket{1_{\phi(n)}}_{[j}\otimes\ket{\psi_{j}(01_{\phi(n)})}
+	\ket{1_{\xi(n)}}_{[j}\otimes \ket{1_{\phi}}_{[j}\otimes\ket{\psi_{j}(1_{\xi}1_{\phi})}\bigg],
\end{align}
where $\ket{\psi_{j}(0_{\xi(n)}0_{\phi(n)})}$, $\ket{\psi_{j}(1_{\xi(n)}0)}$, $\ket{\psi_{j}(01_{\phi(n)})}$, and $\ket{\psi_{j}(1_{\xi(n)}1_{\phi(n)})}$ are the conditional vectors from  the Hilbert space $\mathcal{H}_{\mathcal{S}}$ which satisfy the set of coupled recurrence equations:
		\begin{align}\label{TH-6rec2}
	\ket{\psi_{j+1}(0_{\xi(n)}0_{\phi(n)})} = &\hat{V}_{\eta_{j+1},00}\ket{\psi_{j}(0_{\xi(n)}0_{\phi(n)})}+\sqrt{\tau}\xi_{j}(n)\hat{V}_{\eta_{j+1},10}\ket{\psi_{j}(1_{\xi(n)}0)}
	\nonumber\\&+\sqrt{\tau}\phi_{j}\hat{V}_{\eta_{j+1},01}\ket{\psi_{j}(01_{\phi(n)})}\nonumber\\&+\tau\xi_{j}(n)\phi_{j}(n)\hat{V}_{\eta_{j+1},11}\ket{\psi_{j}(1_{\xi(n)}1_{\phi(n)})},\\
	\ket{\psi_{j+1}(1_{\xi(n)}0)} =& \hat{V}_{\eta_{j+1},00}\ket{\psi_{j}(1_{\xi(n)}0)}+\sqrt{\tau}\phi_{j}(n)\hat{V}_{\eta_{j+1},01}\ket{\psi_{j}(1_{\xi(n)}1_{\phi(n)})},\\
	\ket{\psi_{j+1}(01_{\phi(n)})} =&\hat{V}_{\eta_{j+1},00}\ket{\psi_{j}(01_{\phi(n)})}+\sqrt{\tau}\xi_{j}(n)\hat{V}_{\eta_{j+1},10}\ket{\psi_{j}(1_{\xi(n)}1_{\phi(n)})},
	\\
	\ket{\psi_{j+1}(1_{\xi(n)}1_{\phi(n)})} =&\hat{V}_{\eta_{j+1},00}\ket{\psi_{j}(1_{\xi(n)}1_{\phi(n)})}\label{TH-6rec3}
	\end{align}
and initially $\ket{\psi_{j=0}(0_{\xi(n)}0_{\phi(n)})}=\ket{\psi_{j=0}(1_{\xi(n)}0)}=\ket{\psi_{j=0}(01_{\phi(n)})}=0$, $\ket{\psi_{j=0}(1_{\xi(n)}1_{\phi(n)})}=\ket{\psi_{0}}$. 
  \end{Theorem}

\noindent The proof is analogous to the proof of Theorem 2. 

\section{Proof to Theorem \ref{TH-3}}\label{ProfTH-3}

By applying the standard inner product in  $L^2\left([t_0,\infty)^2\right)$, the formula (\ref{tpa2}) for the two-photon absorption probability can be written in the form
\begin{equation}
    P_{f}(t) = \Gamma_e\Gamma_{f} e^{-\Gamma_f t} 
    \left|
    \bra{K_t} I + \mathcal{F} \ket{\Phi}
    \right|^2 = 4\Gamma_e\Gamma_{f} e^{-\Gamma_f t} 
    \left|
    \bra{K_t} \Pi_\mathcal{S} \ket{\Phi}
    \right|^2,
\end{equation}
where $K_t^\ast = e^{\left(i\omega_{fe}+\frac{1}{2}(\Gamma_f-\Gamma_e)\right)t_2}e^{\left(i\omega_{eg}+\frac{\Gamma_e}{2}\right)t_1} \chi_{t_0<t_1<t_2<t}$ and $\Phi$ are the element of $L^2\left([t_0,\infty)^2\right)$.
In order to maximize the above expression, we have to maximize
\begin{equation}
\left|\bra{K_{t}}\Pi_\mathcal{S}\ket{\Phi}\right|^2,
\end{equation}
The absolute value of the inner product of two vectors is maximized if they are co-linear, hence the optimal $\Phi$ is proportional to $\Pi_\mathcal{S} K_t$. After normalisation,  we obtain
\begin{align}
\Phi_{opt}(t_2,t_1)= \frac{1}{\sqrt{\mathcal{N}_{opt}}}
& \left(e^{\frac{1}{2}(\Gamma_f-\Gamma_e)t_2}e^{\frac{\Gamma_e}{2}t_1} e^{-i\omega_{fe}t_2-i\omega_{eg}t_1}\chi_{t_0<t_1<t_2<t}\right.\nonumber\\
&\left.+e^{\frac{1}{2}(\Gamma_f-\Gamma_e)t_1}e^{\frac{\Gamma_e}{2}t_2}e^{-i\omega_{fe}t_1-i\omega_{eg}t_2}\chi_{t_0<t_2<t_1<t}\right),
\end{align}
where $\mathcal{N}_{opt}$ is given by (\ref{normopt1}) and $\chi_{t_0<t_1<t_2<t}$ is the characteristic function defined on the $\mathbb{R}^2$ plane for the variables $(t_2,t_1)$. 

\section{Proof to Theorem \ref{TH-4}}\label{ProfTH-4}

The formula (\ref{tpa4}) for the two-photon absorption probability can be rewritten  using the the standard inner product in the space $L^{2}\left([t_0,+\infty)^2\right)$ as 
\begin{equation}\label{}
P_{f}(t)  = \Gamma_e\Gamma_{f} e^{-\Gamma_f t} \left|
\braket{K_t}{\Phi}
\right|^2,
\end{equation}
where
\begin{equation}
    K_{t}^{\ast}(t_2,t_1) = 
    e^{\left(i\omega_{fe}+\frac{1}{2}(\Gamma_f-\Gamma_e)\right)t_2}e^{\left(i\omega_{eg}+\frac{\Gamma_e}{2}\right)t_1} \chi_{t_0<t_1<t_2<t} \in L^2\left([t_0,\infty)^2\right).
\end{equation}
The absolute value of the inner product of two vectors is maximized if they are colinear, hence the optimal $\Phi$ is proportional to $K_t$. 
Hence, after a normalisation, we obtain
\begin{equation}
\Phi_{opt}(t_2,t_1)= \frac{1}{\sqrt{\mathcal{N}_{opt}}}e^{\frac{1}{2}(\Gamma_f-\Gamma_e)t_2}e^{\frac{\Gamma_e}{2}t_1} e^{-i\omega_{fe}t_2-i\omega_{eg}t_1}\chi_{t_0<t_1<t_2<t},
\end{equation}
where $\mathcal{N}_{opt}$ is given by (\ref{normopt}).

\end{document}